\documentclass[a4paper,12pt]{article}
\pdfoutput=1 % if your are submitting a pdflatex (i.e. if you have
\usepackage{jheppub} % for details on the use of the package, please
\usepackage{tabularx,subcaption,fancyhdr,mathtools,physics}
\usepackage[makeroom]{cancel}
\usepackage{bm}
\usepackage{csquotes}
\usepackage[nottoc,notlot,notlof]{tocbibind}
\usepackage{float}
\usepackage{relsize}
\usepackage{hyperref}
\usepackage{color,latexsym, array,multirow,url,verbatim, enumerate,cancel}
\DeclareUnicodeCharacter{2212}{-}
\hypersetup{colorlinks=true,linkcolor=blue,citecolor=magenta,filecolor=magenta,
	urlcolor=cyan}
\def \st1{\widetilde t_1}
\def \mst1{m_{\st1}}
\def \sbot1{\widetilde b_1}

\def \lspone{\widetilde\chi_1^0}
\def \mlspone{m_{\lspone}}

\def\chonepm{\widetilde{\chi}_1^{\pm}}
\def\chonemp{\widetilde{\chi}_1^{\mp}}

\def\mchonepm{m_{\chonepm}}

\newcommand{\beq}{\begin{equation}}
\newcommand{\eeq}{\end{equation}}
\def\bea{\begin{eqnarray}}
\def\eea{\end{eqnarray}}

\def \met{\rm E{\!\!\!/}_T}

\title{Complementary probes of bilinear R-parity violating SUSY models with a winolike LSP via neutrino oscillations and at the LHC}

\author[1,*]{Arghya Choudhury,\note[*]{arghya@iitp.ac.in}} 

\author[1,2,\dagger]{Arpita Mondal\note[$\dagger$]{arpita.mondal@lpsc.in2p3.fr (Corresponding author)}}

\affiliation[1]{Department of Physics, Indian Institute of Technology Patna, Bihar - 801106, India}

\affiliation[2]{Laboratoire de Physique Subatomique et de Cosmologie (LPSC), Universit\'e Grenoble-Alpes, CNRS/IN2P3, 53 Avenue des Martyrs, F-38026 Grenoble, France}

\abstract{In this work, we explore the bilinear R-parity violating Supersymmetry model's parameter space by performing a Markov Chain Monte Carlo scan with neutrino oscillation data, Higgs mass and its coupling strengths, and flavor observables such as $B$-hadron decay branching ratios. From the allowed parameter space, we analyze the decay patterns of wino-like lighter charginos and lightest neutralinos and demonstrate how the branching ratios to different neutrino and charged lepton flavors depend on the neutrino mass hierarchy. 
Furthermore, we investigate the impact of current LHC bounds and projected future sensitivities from trilepton resonance searches on the allowed parameter space. We show that considering the branching ratio $\mathrm{Br}(\widetilde{\chi}_1^{\pm} \to Zl^\pm; l= e,\mu,\tau) \sim$23\%, obtained at the best-fit point, 
the wino-like mass degenerate $\widetilde{\chi}_1^{\pm}/\widetilde{\chi}_1^0$ are excluded upto 565 GeV from LHC Run-II  data. The projected exclusion reach with a similar branching ratio at High-Luminosity LHC (HL-LHC) is around 950 GeV. For a simplified scenario where $\widetilde{\chi}_1^{\pm} / \widetilde{\chi}_1^0$ decays via a $Z$ boson with branching ratios of 1\%, 50\%, and 100\%, wino masses can be excluded up to approximately $600~\mathrm{GeV}$, $1185~\mathrm{GeV}$, and $1350~\mathrm{GeV}$ respectively.
Our analysis shows that the HL-LHC can probe a significant portion of the 1$\sigma$ allowed parameter space  by neutrino oscillation measurements and other experimental constraints.}

\begin{document} 
\maketitle

\section{Introduction}
\label{sec:intro}
The discovery of neutrino oscillations~\cite{Borexino:2013zhu,KamLAND:2013rgu,RENO:2018dro,DayaBay:2018yms,Super-Kamiokande:2019gzr,T2K:2018rhz,NOvA:2019cyt} has undoubtedly established that neutrinos have mass and undergo flavor mixing. These findings present a fundamental challenge to the Standard Model (SM)~\cite{PhysRevLett.19.1264,Salam:1968rm,GELLMANN1964214}, which assumes neutrinos to be massless. Understanding the origin and smallness of neutrino masses, as well as their mixing patterns, remains one of the central questions and a key driver for exploring physics beyond the Standard Model (BSM). Supersymmetry (SUSY)~\cite{Drees:2004jm, Baer:2006rs, Martin:1997ns} provides an elegant solution to several longstanding issues in particle physics, such as the hierarchy problem~\cite{SUSSKIND1984181,PhysRevD.14.1667} and gauge coupling unification~\cite{ELLIS1990441,AMALDI1991447,ROSS1992571}, existence of Dark Matter (DM)~\cite{Zwicky:1933gu,1937ApJ....86..217Z,Sofue:2000jx,Jungman:1995df} in the form of lightest supersymmetric particle (LSP), etc. Still, the Minimal Supersymmetric Standard Model (MSSM) cannot explain the existence of non-zero neutrino masses and their mixing. But it can naturally offer a mechanism to generate neutrino masses when extended to include R-parity violation (RPV)~\cite{Dreiner:1997uz, Barbier:2004ez, Choudhury:2024ggy, Banks:1995by,Grossman:1998py,Davidson:2000uc,Davidson:2000ne,Borzumati:1996hd,Mukhopadhyaya:1998xj,PhysRevD.59.091701,PhysRevD.61.055006,Allanach:2007qc,Allanach:2011de,Grossman:1997is,Dreiner:1991pe,Dercks:2017lfq,Bose:2014vea,Datta:2009dc,Das:2005mr,Mitsou:2015kpa,Cohen:2019cge}. Neutrino masses and mixing are commonly generated by seesaw mechanisms, which implement the dimension-5 Weinberg operator~\cite{Weinberg:1979sa, Weinberg:1980bf} by extending the SM with particles like singlet fermions (Type-I), scalar triplets (Type-II), or fermionic triplets (Type-III)~\cite{Minkowski:1977sc,Gell-Mann:1979vob,Schechter:1980gr,Mohapatra:1979ia,Schechter:1981cv}. Alternatively, RPV MSSM provides a framework to explain neutrino oscillations without invoking the Weinberg operator, although the operator can still be generated from RPV terms at low energies. 

The superpotential of the RPV MSSM~\cite{Dreiner:1997uz, Barbier:2004ez, Choudhury:2024ggy} is given by 
%%%%%%%%%%%%%%%%%%%%%%5
\begin{equation}
\label{eq:rpv_potential}
W_{\cancel{R}_p} = \epsilon_i \hat{L}_i \hat{H}_u + \frac{1}{2}\lambda_{ijk}\hat{L}_i\hat{L}_j\hat{E}_k^c + \lambda^{\prime}_{ijk}\hat{L}_i\hat{Q}_j\hat{D}_k^c + \frac{1}{2}\lambda^{\prime\prime}_{ijk}\hat{U}_i^c\hat{U}_j^c\hat{D}_k^c. 
\end{equation}
%%%%%%%%%%%%%%%%%%%%%%%%%%
Here, the first three terms refer to the lepton number violation, and the last term correspond to the baryon number violation. $\hat{L}_i$ ($\hat{E}_k$) represents the left-handed (right-handed) lepton supermultiplet, and $\hat{H}_u$ corresponds to the up-type Higgs supermultiplet. Likewise,  $\hat{Q}_j$, $\hat{U}_j$, and $\hat{D}_k$ denote the left-handed quark doublet, the right-handed up-type quark singlet, and the right-handed down-type quark singlet supermultiplets, respectively. $\epsilon_i$ refers to the bilinear coupling, which shows the coupling between the neutrino and the Higgsinos. The trilinear coupling $\lambda_{ijk}$ ($\lambda_{ijk}^{\prime}$) shows coupling between lepton, slepton, and neutrino (quark, squark, and neutrino). Similarly, the baryon number violating coupling $\lambda^{\prime\prime}_{ijk}$ reflects the coupling between squark and quarks. All three lepton number violating terms  contribute to the generation of the neutrino masses. However, only the bilinear RPV (\texttt{bRPV}) term can generate neutrino mass at both the tree-level and loop-level~\cite{Grossman:1997is, Rakshit:2004rj, Grossman:2003gq}. The other two terms, associated with $\lambda$ and $\lambda^{\prime}$, couplings generate neutrino masses exclusively at the loop-level\footnote{In a recent work~\cite{Choudhury:2024yxd}, authors have explicitly explored the parameter space of trilinear lepton number violating model in the context of neutrino oscillation data for bino and stop LSP and shown the possible collider signatures at the Large Hadron Collider (LHC).}. Bilinear RPV term can exist independently or be generated from trilinear terms via renormalization group evolution, and vice versa~\cite{Roy:1996bua,deCarlos:1996ecd,Nardi:1996iy}. So, in this work, we focus on the \texttt{bRPV} model to explain the neutrino oscillation data.
 
There are a few existing studies in the literature that explore neutrino data with \texttt{bRPV} model~\cite{Hempfling:1995wj,Hirsch:2000ef,Hundi:2011si,Diaz:2014jta,Hirsch:2000jt,Abada:2001zh,Diaz:2004fu,deCampos:2012pf,Gozdz:2008zz,Choudhury:2023lbp,Dreiner:2022zsc}. In a recent study~\cite{Choudhury:2023lbp}, the parameter space of the \texttt{bRPV} model was examined in the context of neutrino oscillation data, focusing on a bino-type neutralino LSP and a wino-type chargino NLSP. In this work, however, we consider both the lightest neutralino and lighter chargino to be wino-type, where both decay directly into different bosons ($W^{\pm}$, $Z$, and $h$) and leptons or neutrinos~\cite{FileviezPerez:2012mj,Dumitru:2018jyb,Dumitru:2018nct}. Such signals have been investigated by the ATLAS Collaboration at the LHC using Run-II data~\cite{ATLAS:2020uer}, providing limits on wino-type lighter chargino ($\chonepm$) or lightest neutralino ($\lspone$) masses based on the branching ratio to $Z$ boson. This branching fraction to different flavors of neutrinos or leptons highly depends on the neutrino mass hierarchy~\cite{FileviezPerez:2012mj,Dumitru:2018jyb,Dumitru:2018nct}. So, we explore this scenario, extract the allowed parameter space, and study the impact of neutrino data on the decay branching ratios of charginos and neutralinos. 
In addition to the neutrino oscillation data, we also consider Higgs data, including the Higgs boson mass and its coupling strengths with various SM particles. Furthermore, flavor physics observables, such as $b$-hadron decay branching ratios, are included in our analysis. Given the large number of parameters in the \texttt{bRPV} model and the extensive set of observables considered, we employ the Markov Chain Monte Carlo (MCMC) method to thoroughly scan the parameter space and identify the allowed regions that satisfy all constraints. Additionally, during the parameter space scanning, we incorporate the latest exclusion limits on sparticle masses.

% Choudhury:2024yxd, Choudhury:2023lbp
% Dreiner:2023bvs,Choudhury:2024ggy,Dreiner:2025kfd

The strongly interacting (colored) sparticles, namely the squarks and gluinos, are excluded up to masses of a few TeV by the current LHC data~\cite{atlas_web,cms_web}. In contrast, the constraints on the electroweakino and slepton sectors are comparatively weaker in both R-parity conserving (RPC)  and R-parity violating (RPV)  scenarios~\cite{atlas_web,cms_web}. 
In RPC scenarios, light electroweakinos and sleptons play a crucial role in dark matter phenomenology, and several phenomenological analyses have addressed this issue along with 
the interpretations of LHC limits for various production and decay modes \cite{Wang:2026owv, Constantin:2025bqp, Chatterjee:2025gej, Choudhury:2024crp, Cornell:2024dki, Altakach:2023tsd, Chakraborti:2017dpu,  Bisal:2024ezn, Chowdhury:2016qnz, Barman:2016kgt, Choudhury:2016lku, Chakraborti:2015mra, Chakraborti:2014gea, Choudhury:2013jpa, Choudhury:2012tc, Bhattacharyya:2011se}. 
These light sparticles in the RPV scenarios have more diverse final states depending on the non-zero RPV couplings and the phenomenological implications 
have been studied in refs.\cite{Barman:2026cez, Cottin:2025avd, Baruah:2024wrn, Dreiner:2025kfd, Choudhury:2023yfg, Bhattacherjee:2023kxw, Dreiner:2023bvs, Choudhury:2023eje, Barman:2020azo, Barman:2025bpx, Choudhury:2024ggy}. 
However the collider phenomenology with RPV SUSY models is relatively less explored. 
It is worth mentioning that the electroweakinos, particularly the charginos and neutralinos, play a direct role in neutrino mass generation in the present framework. 
Motivated by these considerations, we also explore the collider implications of the model in the context of trilepton resonance search at the Large Hadron Collider (LHC). Using the Run-II results of trilepton resonance search provided by the ATLAS Collaboration~\cite{ATLAS:2020uer} with an integrated luminosity $\mathcal{L}$ = 139 fb$^{-1}$, we reinterpret the exclusion limits on wino-like $\chonepm$ or $\lspone$ mass as a function of its branching ratio to the $Z$ boson. We compare these limits related to its branching ratios predicted from the neutrino sector and show that the parameter space allowed by neutrino oscillation data, along with other data, remains consistent with current LHC constraints. Furthermore, we estimate the future sensitivity of the High-Luminosity LHC (HL-LHC) with $\mathcal{L}=3$ ab$^{-1}$ and determine the projected exclusion reach for the chargino/neutralino mass as a function of its decay branching ratio to the $Z$ boson.

The paper is organized as follows. In Section~\ref{sec:model}, we discuss the generation of neutrino masses and mixing angles in the \texttt{bRPV} model, along with the observables considered in our analysis. The recent collider limits on sparticles, the parameter space, and the details of the MCMC analysis are presented in Section~\ref{sec:param}. Our results are shown and explained in Section~\ref{sec:result}. Finally, we conclude in Section~\ref{sec:conclusion}.

\section{Neutrino mass generation and observable}
\label{sec:model}
\begin{figure}[!htb]
\begin{center}
    \includegraphics[width=0.3\textwidth]{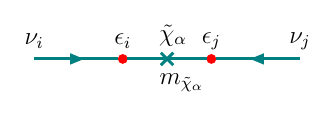}
    \includegraphics[width=0.3\textwidth]{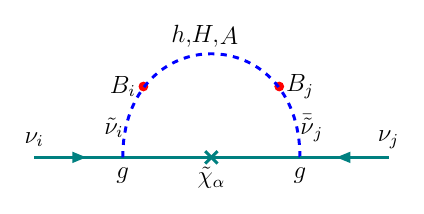}
    \includegraphics[width=0.3\textwidth]{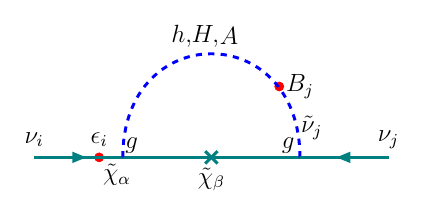}
    \caption{Neutrino mass generation mechanisms: tree-level (left), $BB$-loop (middle), and $\epsilon B$-loop (right)~\cite{Choudhury:2023lbp}.}
    \label{fig:feynman_diagram}
\end{center}
\end{figure}

The lagrangian from the superpotential of the \texttt{bRPV} and the soft term in the lagrangian~\cite{Dreiner:1997uz, Barbier:2004ez, Banks:1995by} can be written as 
\begin{equation}
\mathcal{L} = \epsilon_i \Big(\tilde{H}_u^0 \nu_{iL} - \tilde{H}_u^+ l_{iL}\Big) + \text{h.c}; ~~~~~~~
\mathcal{L}_{\text{soft}} = B_i \Big( \tilde{\nu}_i H_u^0 - \tilde{l}_i^- H_u^+ \Big) + \text{h.c.},
\end{equation}
where $\epsilon_i$ represents the coupling between neutrino ($\nu_i$) and Higgsino ($\tilde{H}_u^0$) and lepton ($l_i$) and Higgsino ($\tilde{H}_u^+$). Similarly, the soft coupling $B_i$ refers to the coupling between sneutrino ($\tilde{\nu}$) and Higgs ($H_u^0$) and slepton ($\tilde{l}$) and Higgs ($H_u^+$). Because of the coupling of up-type Higgsinos with neutrinos, neutrino mass can be generated at the tree-level. At this level, the neutralino-neutrino mass matrix becomes a $7 \times 7$ structure in the basis~\cite{Banks:1995by, Nardi:1996iy, Grossman:1998py, Rakshit:2004rj,Choudhury:2023lbp} $\left( \begin{matrix} 
\tilde{B} & \tilde{W}_3 & \tilde{H}_d^0 & \tilde{H}_u^0 & \nu_e & \nu_{\mu} & \nu_{\tau} 
\end{matrix} \right)$, where $\tilde{B}$ and $\tilde{W}_3$ refer to the bino and wino states, respectively. Additionally, couplings between different Higgs states ($h, H, A$) and sneutrinos lead to sneutrino mass splitting, which contributes to neutrino mass generation through loop-level processes~\cite{Grossman:1997is}. These mechanisms are illustrated in Figure~\ref{fig:feynman_diagram}, showcasing neutrino mass generation via the tree-level process (left), $BB$-loop process (middle), and $\epsilon B$-loop process (right).

%%%%%%%%%%%%%%%%%%%%%%%%%%%%%%%%%%%%%%%%%%%% 
%%%%%%%%%%%%%%%%%%%%%%%%%%%%%%%%%%%%%%%%%%%%%%%
Combining these three contributions one can write the neutrino mass  as \cite{Rakshit:2004rj,Barbier:2004ez}
\begin{equation} 
\begin{split}
\label{eq:total_mass}
[m_{\nu}]_{ij}  = X_T \epsilon_i \epsilon_j \sin^2\zeta + C_{ij} B_iB_j + (C_{ij}^\prime \epsilon_i B_j + i \leftrightarrow j), 
\end{split}
\end{equation}
where first, second, and third term represent the tree-level, $BB$-loop, and $\epsilon B$-loop contribution respectively. $\zeta$ shows the alignment between $\epsilon_i$ parameters and sneutrino vev ($v_i$)~\cite{Chun:2002vp,Grossman:2000ex,Nardi:1996iy,Borzumati:1996hd}. The definition of $X_T$ is given by \cite{Rakshit:2004rj,Grossman:2003gq}
\beq
\label{eq:XT}
X_T = {m_Z^2 m_{\tilde \gamma}\cos^2\beta \over 
\mu(m_Z^2 m_{\tilde \gamma}\sin 2\beta-M_1 M_2 \mu)}
\eeq 
where $m_{\tilde \gamma}\equiv \cos^2\theta_w M_1 + \sin^2\theta_w M_2$.
Assuming all the neutrino masses at the EWSB scale ($\tilde{m}$), the approximate neutrino masses at the tree-level and from the two loop-level processes are expressed as~\cite{Grossman:2003gq,Choudhury:2023lbp,Choudhury:2024ggy}:  
\begin{align}
\label{eq:tree}
[m_{\nu}]_{ij}^{\epsilon\epsilon} & \sim \frac{\cos^2\beta}{\tilde m} \epsilon_i \epsilon_j \sin^2\zeta, \\
\label{eq:BB}
[m_{\nu}]_{ij}^{BB} & \sim \frac{g^2}{64\pi^2\cos^2\beta} \frac{B_iB_j}{\tilde{m}^3} \, \epsilon_H, \\
\label{eq:epsilonB}
[m_{\nu}]_{ij}^{\epsilon B} & \sim \frac{g^2}{64\pi^2\cos\beta} \frac{\epsilon_iB_j + \epsilon_jB_i}{\tilde{m}^2} \, \epsilon_H^\prime,
\end{align}
where $\tilde{m}$ represents the electroweak symmetry breaking scale, and $\sin^2\zeta$ corresponds to the alignment between the \texttt{bRPV} coupling ($\epsilon$) and the sneutrino vacuum expectation value ($v$)~\cite{Chun:2002vp,Grossman:2000ex,Nardi:1996iy,Borzumati:1996hd}. The factors $\epsilon_H$ and $\epsilon_H^\prime$ quantify the cancellation effects of different Higgs states in the $BB$-loop and $\epsilon B$-loop, respectively~\cite{Grossman:2003gq,Rakshit:2004rj,Davidson:2000uc}. It is noteworthy that at the tree-level, only a single neutrino, the heaviest mass eigenstate, acquires a significant mass. 

Now using this \texttt{bRPV} model, we attempt to satisfy the neutrino oscillation parameters~\cite{deSalas:2020pgw} corresponding to both neutrino mass orderings: the Normal Hierarchy (NH) with $m_{\nu_3}> m_{\nu_2} > m_{\nu_1}$ and Inverted Hierarchy (IH) with $m_{\nu_2}> m_{\nu_1} > m_{\nu_3}$. Each scenario includes two mass square splittings ($\Delta m_{21}^2$ and $|\Delta m_{31}^2|$) and three mixing angles ($\theta_{13}$, $\theta_{12}$, and $\theta_{23}$). It should be noted that we have not considered the CP-violating phase ($\delta_{\text{CP}}$) since adding this phase does not alter the parameter space~\cite{Choudhury:2023lbp}. We have considered the constraint on the neutrino masses sum coming from the cosmological data~\cite{deSalas:2020pgw}. 
Additionally, we incorporate the Higgs data, such as the Higgs boson mass~\cite{Allanach:2004rh} and its coupling with various Standard Model particles like $Z,~W,~b,~\tau,~\mu,~t,~\gamma$ given by CMS Collaboration~\cite{cms_web1}. Furthermore, we take into account the branching ratios of $b$-hadron decays, including $\mathcal{B}r(B_s \rightarrow \mu^+ \mu^-)$~\cite{LHCb:2021vsc} and $\mathcal{B}r(B \rightarrow X_s \gamma)$~\cite{HFLAV:2019otj}. The values of these observables, along with the prior on neutrino masses sum under consideration, are listed in Table~\ref{tab:all_obs}.
%%%%%%%%%%%%%%%%%%%%%%%%%%%%%%%%%%%%%%%%%
\begin{table}[!htb]
\centering
\renewcommand{\arraystretch}{1.5} % Adjust the row spacing
\resizebox{\textwidth}{!}{
\begin{tabular}{|c|c|c||c|c||c|c|} 
    \hline
    \multirow{2}{*}{Observable} & \multicolumn{2}{c||}{Best-fit$^{+1\sigma}_{-1\sigma}$} & \multirow{2}{*}{Observable} & \multirow{2}{*}{Best-fit$^{+1\sigma}_{-1\sigma}$} & \multirow{2}{*}{Observable} & \multirow{2}{*}{Best-fit$^{+1\sigma}_{-1\sigma}$} \\
	%\hline
	\cline{2-3}
	& NH & IH & & & & \\
	\hline
	$\Delta m^2_{21}$ [$10^{-5}$ eV$^2$] & 7.50$^{+0.22}_{-0.20}$  & 7.50$^{+0.22}_{-0.20}$  & $m_h$~[GeV] & 125$^{+3}_{-3}$ & $\kappa_{\mu}$ & 0.92$^{+0.55}_{-0.87}$ \\
	\hline
	$|\Delta m^2_{31}|$ [$10^{-3}$ eV$^2$] & 2.55$^{+0.02}_{-0.03}$ & 2.45$^{+0.02}_{-0.03}$ & $\kappa_Z$ & 0.96$^{+0.07}_{-0.07}$ & $\kappa_{t}$ & 1.01$^{+0.11}_{-0.11}$  \\
	\hline
	$\theta_{12}$ [$^{\circ}$] & 34.3$^{+1.0}_{-1.0}$ & 34.3$^{+1.0}_{-1.0}$ & $\kappa_{W}$ & 1.11$^{+0.14}_{-0.09}$  & $\kappa_{\gamma}$ & 1.01$^{+0.09}_{-0.14}$  \\
	\hline
	$\theta_{13}$ [$^{\circ}$] & 8.53$^{+0.13}_{-0.12}$ & 8.58$^{+0.12}_{-0.14}$  &$\kappa_{b}$ & 1.18$^{+0.19}_{-0.27}$  & $\mathcal{B}r(B_s \rightarrow \mu^+ \mu^-)~[10^{-9}]$ & 3.09$^{+0.48}_{-0.44}$   \\
	\hline\cline{5-6}
	$\theta_{23}$ [$^{\circ}$] & 49.26$^{+0.79}_{-0.79}$ & 49.46$^{+0.60}_{-0.97}$ & $\kappa_{\tau}$ & 0.94$^{+0.12}_{-0.12}$ & $\mathcal{B}r(B \rightarrow X_s \gamma)~[10^{-4}]$ & 3.32$^{+0.15}_{-0.15}$  \\
    \hline\hline
    \multicolumn{3}{|c||}{Prior} & \multicolumn{2}{c||}{NH} & \multicolumn{2}{c|}{IH} \\
    \hline
    \multicolumn{3}{|c||}{Cosmological bound, $\sum_i m_{\nu_i} $} & \multicolumn{2}{c||}{$< 0.12$ eV} & \multicolumn{2}{c|}{$< 0.15$ eV} \\
    \hline
    %\multicolumn{6}{|c|}{$\sum_i m_{\nu_i}$ $\leq 0.12$~eV}\\
    %\hline
\end{tabular}
}
\caption{Neutrino observables~\cite{deSalas:2020pgw} and the cosmological bound on neutrino masses sum~\cite{Planck:2018vyg} corresponding to Normal Hierarchy (NH) and Inverted Hierarchy (IH), Higgs mass~\cite{Allanach:2004rh}, Higgs coupling strength modifiers~\cite{cms_web1}, and $b$-hadron decay branching ratio~\cite{LHCb:2021vsc,HFLAV:2019otj} best-fit values with their $1\sigma$ uncertainties are shown.
}
\label{tab:all_obs}
\end{table}
%%%%%%%%%%%%%%%%%%%%%%%%%%%%%%%%%%%%%%%%%
 
\section{Parameter space and analysis details}
\label{sec:param}
In this analysis, we have considered a \texttt{bRPV} model with wino-type LSP. In this model, the lightest neutralino ($\lspone$) and the lighter chargino ($\chonepm$) are almost mass degenerate. The strongest limits on sparticle masses in the RPV scenarios mainly come from \texttt{LLE} ($\lambda$) type couplings. Using the Run-II data, the ATLAS and CMS Collaborations have excluded gluino, light squarks, stop, slepton, and chargino upto 2.5 TeV~\cite{ATLAS:2021yyr}, 1.6 TeV~\cite{CMS:2016zgb}, 1.9 TeV~\cite{CMS:2013pkf}, 1.2 TeV~\cite{ATLAS:2021yyr}, and 1.6 TeV~\cite{ATLAS:2021yyr} respectively. However, the ATLAS Collaboration has searched for winos and Higgsinos in the \texttt{bRPV} scenario. Higgsinos in the \texttt{bRPV} model are excluded upto 440 GeV~\cite{ATLAS:2023lfr}, and winos are excluded in the mass range 100-1100 GeV~\cite{ATLAS:2017jvy, ATLAS:2024zkx} depending on the assumptions of decay branching ratio into different flavors of leptons. The decay channels corresponding to different LSPs for different RPV coupling are discussed in details in the Ref.~\cite {Dreiner:2023bvs,Choudhury:2024ggy,Dreiner:2025kfd} and all the limits on these SUSY particles coming from RPV couplings are summarized in the Refs.~\cite{atlas_web, cms_web}.

%For this scenario, the lower bounds on $\mlspone$ and $\mchonepm$ depend on their decay branching ratio to $Z$ boson as shown in the Ref.~\cite{ATLAS:2020uer}. Among other RPV couplings, the strongest limit on $\mchonepm$ is coming from the $\lambda$ or \texttt{LLE} type coupling and that $\sim$1.6 TeV~\cite{ATLAS:2021yyr}. Similarly, the slepton masses are excluded upto $\sim$1.2 TeV with \texttt{LLE} type coupling~\cite{ATLAS:2021yyr}. Similarly, the ATLAS Collaboration has excluded gluino masses upto $\sim$2.5 TeV for \texttt{LLE} type coupling~\cite{ATLAS:2021yyr}. The exclusion limits on the masses of stop NLSP ranges from $\sim$820 to 1020 GeV, depending on different RPV couplings~\cite{CMS:2013pkf}. For degenerate light squark NLSPs, the limit is set at $\sim$1.6 TeV~\cite{CMS:2016zgb}. The ATLAS experiment excludes $\tilde{t}_1$ LSP masses up to 1.9 TeV, 1.8 TeV, and 800 GeV for decays to $be$, $b\mu$, and $b\tau$, respectively, assuming a 100\% branching ratio~\cite{ATLAS:2017jvy, ATLAS:2024zkx}. The decay channels corresponding to different LSP for different RPV coupling are discussed in details in the Ref.~\cite{Dreiner:2023bvs,Choudhury:2024ggy,Dreiner:2025kfd} and all the limits on these SUSY particles coming from RPV couplings are summarized in the Ref.~\cite{atlas_web, cms_web}   
%%%%%%%%%%%%%%%%%%%%%%%%%%%%%%%%%%%%%%%%%
\begin{table}[!htb]
\centering
\resizebox{\textwidth}{!}{
\begin{tabular}{|c|c||c|c||c|c|} 
    \hline
    Parameter & Range/Value & Parameter & Range/Value & Parameter & Range/Value \\
	\hline
	$M_1$ [GeV] & 3000 & $\tan\beta$ & 1-60 & $B_i$[GeV] & 1-5000 \\
	\hline
	$M_2$ [GeV] & 500-2000 & $M_A$ [GeV] & 3000 & $A_t$ [GeV] & -3.5 \\
    \hline
    $M_3$ [GeV] & 3000 & $v_i$ [$10^{-4}$GeV] & 0.1-50 & $m_{\tilde{q}}$ [GeV] & 3000 \\
    \hline
    $\mu$ [GeV] & 3000 & $|\epsilon_i|$ [GeV] & 0-0.5 & $m_{\tilde{l}}$ [GeV] & 2000 \\
    \hline
\end{tabular}
}
\caption{The values of fixed parameters and the range of free parameters are shown here.
}
\label{tab:param}
\end{table}
%%%%%%%%%%%%%%%%%%%%%%%%%%%%%%%%%%%%%%%%%
%Heavy Higgs searches via $H/A \rightarrow \tau^+ \tau^-$ set the most stringent limits, excluding $M_A$ up to 1.0 TeV for $\tan\beta < 8$ and 1.5 TeV for $\tan\beta < 21$~\cite{CMS:2018rmh, ATLAS:2020zms}. 
To avoid the existing limits, we have fixed gluino and all the squark masses at 3 TeV and the slepton masses at 2 TeV. Also, we have set $M_A$  and $A_t$ at 3 TeV and -3.5 TeV respectively\footnote{It may be noted that the most strongest limit on $M_A > 1.5$~TeV (for $\tan\beta < 21)$~\cite{CMS:2018rmh, ATLAS:2020zms} is obtained from $H/A \rightarrow \tau^+ \tau^-$ search.}. Also, we have set $M_1$ and $\mu$ both at 3 TeV. Now in this scenario, after fixing these parameters, we have considered 11 free parameters to vary - $M_2$, $\tan\beta$, three sneutrino vev ($v_i$), three \texttt{bRPV} coupling ($\epsilon_i$), and three soft \texttt{bRPV} coupling ($B_i$). The ranges of these input parameters, along with the values of fixed parameters, are mentioned in the Table~\ref{tab:param}.

We have generated our model using the package \texttt{SARAH}~\cite{Staub:2008uz,Staub:2010jh,Staub:2015kfa}, and to generate the spectrum, we have utilized \texttt{SPheno}~\cite{Porod:2003um,Porod:2011nf} which uses \texttt{FlavorKit}\cite{Porod:2014xia} for the calculation of flavor physics observables. 
To scan the parameter space, we employ the Markov Chain Monte Carlo (MCMC) method using the \texttt{emcee} package~\cite{foreman2013emcee}. The MCMC algorithm samples the parameter space according to the likelihood function $\mathcal{L} \propto e^{-\frac{\chi^2}{2}}$
thereby identifying both the maximum-likelihood (minimum $\chi^2$) point and the surrounding high-likelihood regions. The $\chi^2$ function is defined as $\chi^2 = \sum_{i=1}^{n_{\rm obs}} \left[ \frac{X_i^{\rm obs} - X_i^{\rm th}}{\sigma_i} \right]^2$, where $X_i^{\rm obs}$, $X_i^{\rm th}$, and $\sigma_i$ denote the experimentally observed value, theoretically predicted value, and experimental uncertainty of the $i$-th observable, respectively.
%To scan the parameter space, we employ the MCMC method using \texttt{emcee} package~\cite{foreman2013emcee}. Using this MCMC setup, we try to find the maximum likelihood function $\mathcal{L} \propto e^{\frac{\chi^2}{2}}$ i.e, we need to find out the minimum $\chi^2$ value. The $\chi^2$ is defined as $\chi^2 = \sum_{i=1}^{n_{\rm obs}} \left[ \frac{X_i^{\rm obs} - X_i^{\rm th}}{\sigma_i} \right]^2$, where $X_i^{\rm obs}$, $X_i^{\rm th}$, and $\sigma_i$ refer to the experimentally observed value, theoretically calculated value and the experimentally measured uncertainty of each observable. 
In this analysis, we have 15 observables and 11 free parameters, which lead to 4 degrees of freedom (\texttt{d.o.f}). To scan the parameter space thoroughly, we use 500 walkers and 400 steps for each walker. 
%Also, we run it in 12 cores, which leads to $500\times 400\times 12$ total number of generated samples. 
To ensure that it does not depend on the initial steps of the sampling, we employ 30\% burn-in for our analysis.

\section{Results and Discussions}
\label{sec:result}
We present results for both the NH and IH scenarios, while the collider prospects are discussed only for the NH case. The results for the IH scenario are briefly summarized in Section~\ref{sec:ih} and compared with those of the NH case.
%%%%%%%%%%%%%%%%%%%%%%%%%%%
\subsection{Normal Hierarchy}
\label{sec:nh}
\begin{figure}[!htb]
\includegraphics[width=1\textwidth]{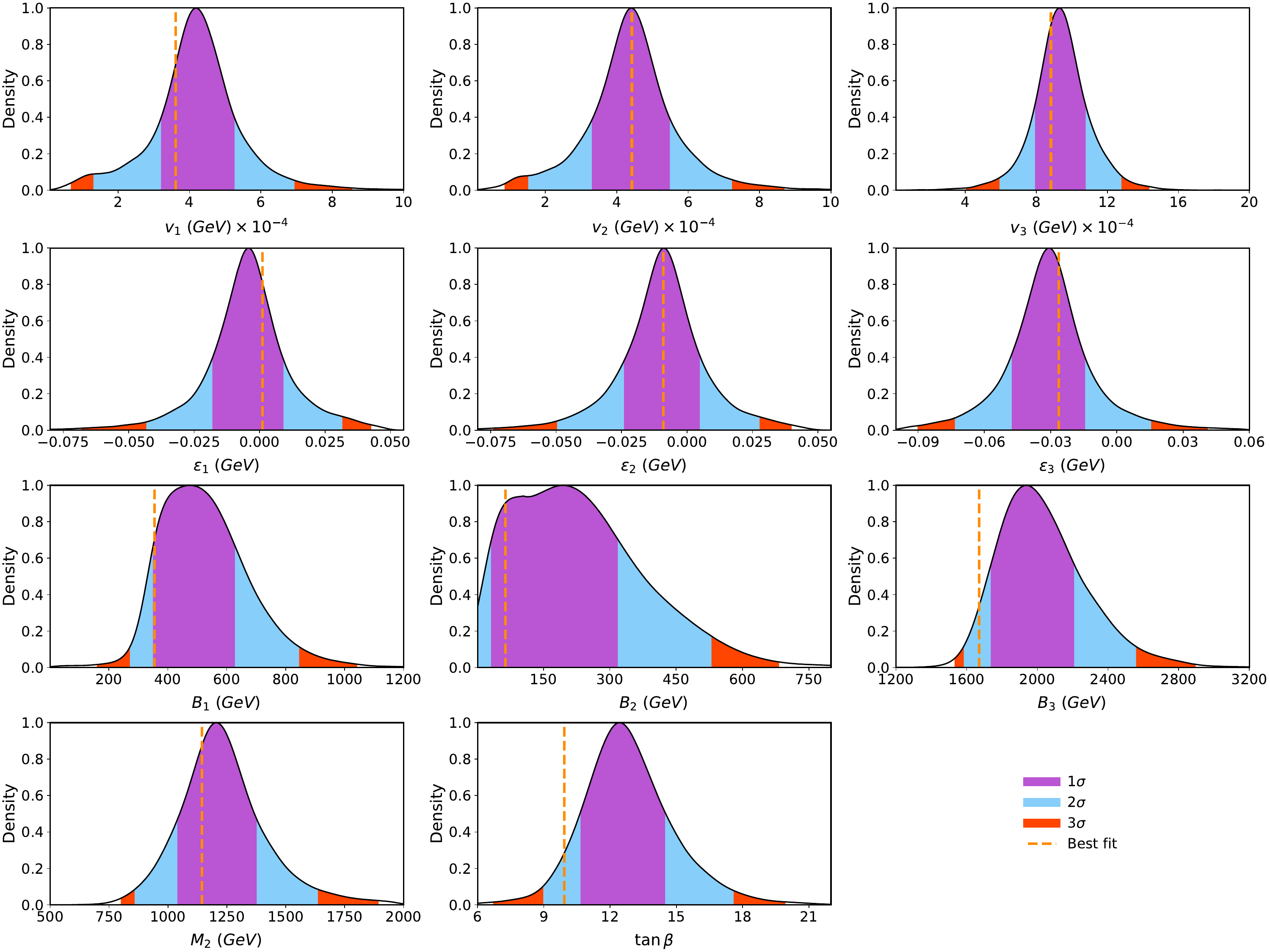}
    \caption{The 1-D posterior distribution of all the input parameter along with $1\sigma$, $2\sigma$ and $3\sigma$ regions are shown here with purple, cyan and red colors respectively. The best-fit value of each parameter is represented by the dashed orange colored line.}
    \label{fig:param_region}
\end{figure}
%%%%%%%%%%%%%%%%%%%%%%%%%

In this section, we consider the Normal Hierarchy scenario, where the third neutrino ($\nu_3$) is the heaviest and the first neutrino ($\nu_1$) is the lightest. In this framework, the minimum $\chi^2$ value is obtained as 3.20 for \texttt{d.o.f} = 4. The posterior distributions of the input parameters are shown in Figure~\ref{fig:param_region}, where the purple, cyan, and red regions correspond to the $1\sigma$, $2\sigma$, and $3\sigma$ confidence intervals, respectively. The best-fit point is indicated by the orange dashed line. The input parameters corresponding to this best-fit point, along with the masses of $\chonepm$ and $\lspone$ and their decay branching ratios, are shown in Table~\ref{tab:best_fit}.
%%%%%%%%%%%%%%%%%%%%%%%%%%%%%%%%%%%%%%%%%
\begin{table}[!htb]
\centering
\resizebox{\textwidth}{!}{
\begin{tabular}{|c|c|c|c||c|c|c|c|} 
    \hline
    \multicolumn{8}{|c|}{Best-fit Point (BFP)}\\
    \hline
    \multicolumn{4}{|c||}{Input parameters} &  \multicolumn{2}{c|}{$m_{\lspone}$ = 1191 GeV} & \multicolumn{2}{c|}{$m_{\chonepm}$ = 1194 GeV} \\
    \hline
    Parameter & Value & Parameter & Value & Decay & Br (\%) & Decay & Br(\%) \\
	\hline
	$M_2$ [GeV] & 1143.89 & $\epsilon_2$ [$10^{-3}$GeV] & -9.05 & $We$ & 8.68& $W\nu_{\tau}$ & 45.35 \\
	%\hline
	$\tan\beta$ & 9.93 & $\epsilon_3$ [$10^{-3}$GeV] & -26.34  & $W\mu$& 8.61& $he$ & 5.11\\
   % \hline
    $v_1$ [$10^{-4}$GeV] & 3.61 & $B_1$ [GeV] & 355  & $W\tau$& 28.10 & $h\mu$& 4.31\\
    %\hline
    $v_2$ [$10^{-4}$GeV] & 4.42 & $B_2$ [GeV] & 64 & $h\nu_e$ &0.06 & $h\tau$ & 21.92 \\
    %\hline
    $v_3$ [$10^{-4}$GeV] & 8.82 & $B_3$ [GeV] & 1670   & $h\nu_{\mu}$& 0.17& $Ze$ & 4.40\\
    $\epsilon_1$ [$10^{-3}$GeV] & -1.03 &  &  & $h\nu_{\tau}$&31.06 & $Z\mu$& 4.46 \\
    \cline{1-4}
    \multicolumn{4}{|c||}{$\chi^2_{\text{min}}/\texttt{d.o.f} = 3.20/4 = 0.8$} & $Z\nu_{\tau}$ & 23.32 & $Z\tau$& 14.45 \\
    \hline
\end{tabular}
}
\caption{The input parameters at the best-fit point, along with the masses of the wino-type lightest neutralino and lighter chargino and their respective decay branching ratios, are presented.
}
\label{tab:best_fit}
\end{table}
%%%%%%%%%%%%%%%%%%%%%%%%%%%%%%%%%%%%%%%%%
Based on the relationships of tree-level and two loop-level contributions with $\tan\beta$ shown in Equations~\ref{eq:total_mass}-\ref{eq:epsilonB}, we observe that $\tan\beta$ acts as a suppression factor for the tree-level contribution while enhancing the loop-level contributions. Since the heaviest neutrino mass primarily arises from the tree-level contribution and the other two from loop-level contributions, $\tan\beta$ should not be excessively large or small. To get the heaviest neutrino mass from the tree-level, the higher $M_2$ values are expected which is evident from the relation of $X_T$ with $M_2$ in Equation~\ref{eq:XT}. These arguments are reflected at the best-fit point shown in Table~\ref{tab:best_fit}. From Figure~\ref{fig:param_region}, the $3\sigma$ allowed range for $\tan\beta$ and $M_2$ are approximately 7-20 and 800-1900 GeV respectively. As the third neutrino is the heaviest one, $\epsilon_3$ and $v_3$ are expected to have larger values than other $\epsilon_i$ and $v_i$ as shown in Table~\ref{tab:best_fit}. As the $\epsilon B$-loop contribution is significantly suppressed due to the small value of $\epsilon$ compared to the $B_i$ parameters, the $B_1$ parameter must be larger than the $B_2$ parameter to make the lightest neutrino mass eigenstate heavy. This is evident from the best-fit values shown in the Table~\ref{tab:best_fit}. %\tcr{The reason behind the hierarchies of these $\epsilon_i$, $v_i$, and $B_i$ parameters are discussed in detail in~\cite{Choudhury:2023lbp}. should we remove this!!!}
%%%%%%%%%%%%%%%%%%%%%%%%%%%%%%%%%%%%

\begin{figure}[!htb]
\includegraphics[width=1\textwidth]{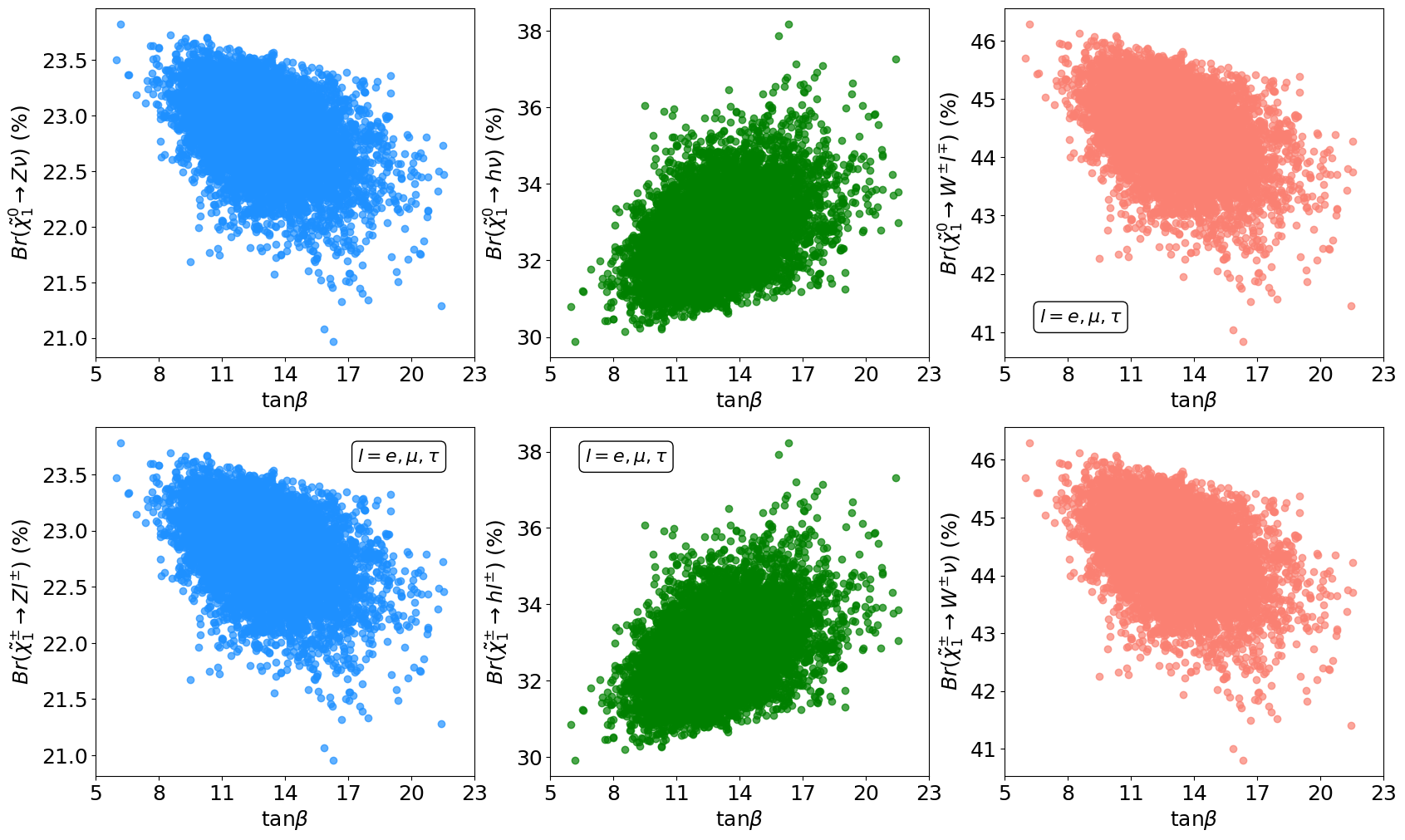}
    \caption{Variations of branching ratios corresponding to the three possible decay channels of a wino-type lightest neutralino ($\lspone$) and a wino-type lighter chargino ($\chonepm$) as functions of $\tan\beta$. The blue, green, and coral points in the upper panel 
 represent $\mathcal{B}r(\lspone \rightarrow Z\nu)$, $\mathcal{B}r(\lspone \rightarrow h\nu)$, and $\mathcal{B}r(\lspone \rightarrow W^{\pm}l^{\mp})$, respectively, with $l = e, \mu, \tau$.  
Similarly, the blue, green, and coral points in the lower panel correspond to  
$\mathcal{B}r(\chonepm \rightarrow Zl^{\pm})$, $\mathcal{B}r(\chonepm \rightarrow hl^{\pm})$, and $\mathcal{B}r(\chonepm \rightarrow W^{\pm}\nu)$, respectively.
%\tcm{use one format either wino or wino !!!!!!!! check also other places... also either write $\lspone$ or \tcb{lightest} neutralino , \tcb{lighter} chargino etc .... Only neutralino/chargino is not correct. Check the main text also. and other places/captions....... Change the next fig caption accordingly as this one.}
}
    \label{fig:br_tanbeta}

\end{figure}
%%%%%%%%%%%%%%%%%%%%%%%%%%%%%%%%%%%%%  
%The neutrino mass hierarchy is reflected in the values of the $\epsilon_i$ and the $v_i$ parameters as presented in Table~\ref{tab:best_fit}. 
 
The ordering % hierarchy in
 of the branching ratios of the $\lspone$ and $\chonepm$ corresponding to different lepton flavors arises from the relative sizes of the parameters which are driven by the neutrino mass hierarchy. Since $m_{\nu_3} > m_{\nu_2} > m_{\nu_1}$, the branching ratios to $\tau$-flavored leptons and neutrinos are the largest.
%%%%%%%%%%%%%%%%%%%%%%%%%%%%%%%
\begin{figure}[!htb]
\includegraphics[width=1\textwidth]{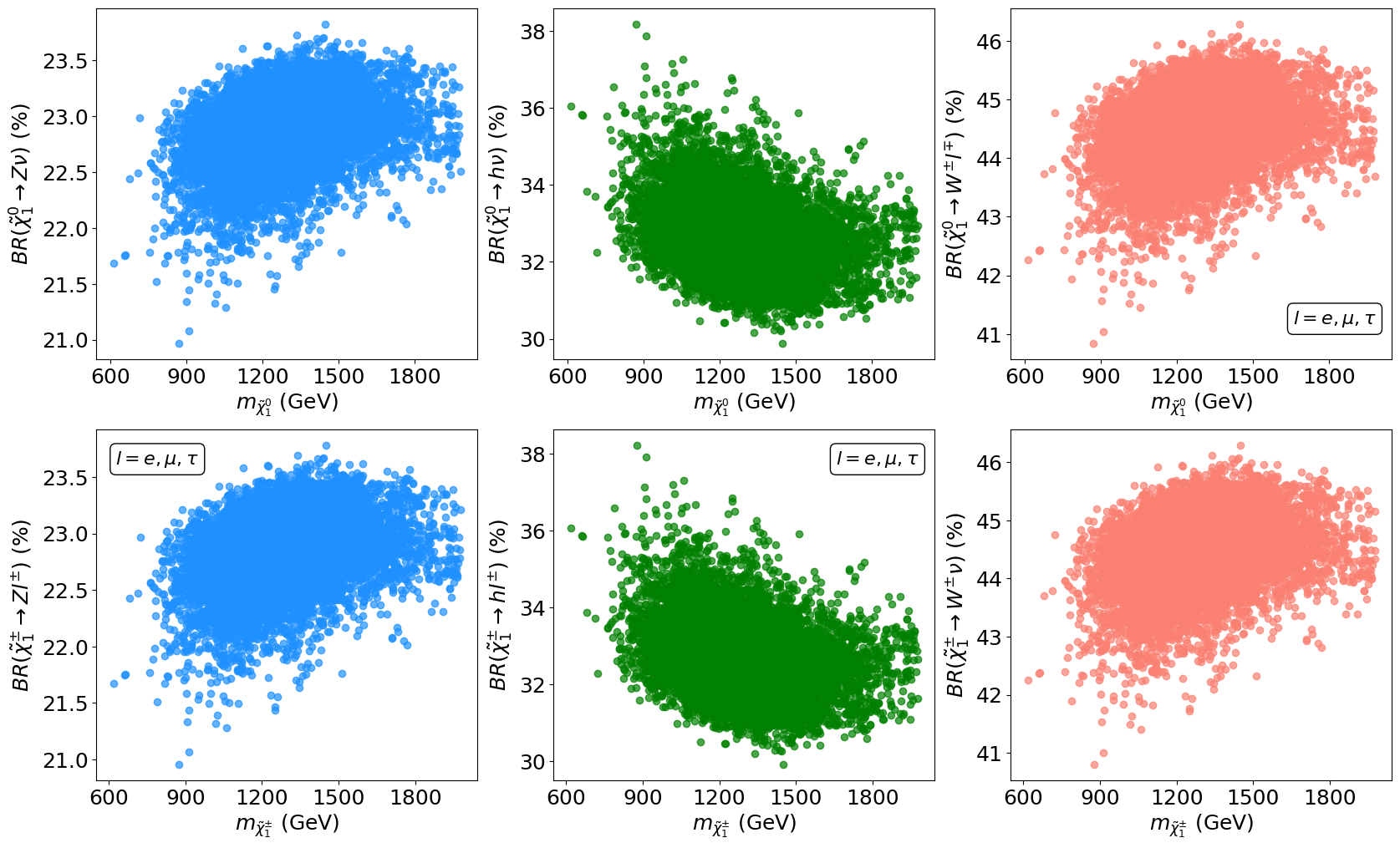}
    \caption{Variations of branching ratios corresponding to the three possible decay channels of $\lspone$ and $\chonepm$ as functions of their masses. The blue, green, and coral points in the upper panel represent $\mathcal{B}r(\lspone \rightarrow Z\nu)$, $\mathcal{B}r(\lspone \rightarrow h\nu)$, and $\mathcal{B}r(\lspone \rightarrow W^{\pm}l^{\mp})$, respectively, with $l = e, \mu, \tau$. Similarly, the blue, green, and coral points in the lower panel represent $\mathcal{B}r(\chonepm \rightarrow Zl^{\pm})$, $\mathcal{B}r(\chonepm \rightarrow hl^{\pm})$, and $\mathcal{B}r(\chonepm \rightarrow W^{\pm}\nu)$, respectively. }
    \label{fig:br_mass}
\end{figure}
%%%%%%%%%%%%%%%%%%%%%%%%%%
Now we discuss the effect of $\tan\beta$ on the branching ratios of the lighter chargino and the lightest neutralino as shown in Figure~\ref{fig:br_tanbeta}. %To do that,  
We consider the points with $\chi^2 \leq \chi^2_{\text{min}} + 9.0$, which is the allowed $\chi^2$ value for a 1D parameter space at the $3\sigma$ level~\cite{ParticleDataGroup:2022pth}. From the large dataset of the allowed points, we randomly select only 5\% points for further illustration purposes. The relations of the branching ratios of wino-type chargino and neutralino are discussed in Ref.~\cite{FileviezPerez:2012mj, Dumitru:2018jyb, Dumitru:2018nct}. The dependency of $\tan\beta$ and the mass of neutralino is shown in Eqs.~(4.3)-(4.5) of Ref.~\cite{Dumitru:2018nct} and Eqs.~(70)-(74) in Ref.~\cite{FileviezPerez:2012mj}. Additionally, the relationship between mass and $\tan\beta$ to the decay branching ratios for the wino-type chargino is shown in Eqs.~(3.7)-(3.9) of Ref.~\cite{Dumitru:2018nct}. 
%The authors in these studies have already discussed that 
For the branching ratios to $Z$ and $W$ bosons
$\tan\beta$ acts as a suppression factor ($\propto 1/\sqrt{1+\tan^2\beta}$), whereas for the Higgs boson, the branching ratio increases with increasing $\tan\beta$ ($\propto \tan^2\beta$) ~\cite{FileviezPerez:2012mj, Dumitru:2018jyb, Dumitru:2018nct}. We have also obtained a similar pattern from our analysis, which is reflected in Figure~\ref{fig:br_tanbeta}. Here we have plotted only the decay modes with a branching fraction $>$ 1\%. In Figure~\ref{fig:br_tanbeta}, $\mathcal{B}r(\lspone \rightarrow Z\nu/h\nu/W^{\pm}l^{\mp})$ and $\mathcal{B}r(\chonepm \rightarrow Zl^{\pm}/hl^{\pm}/W^{\pm}\nu)$ are presented via blue, green, and coral colored points, respectively with $l=e,\mu,\tau$. %In the upper panel we have shown the decay channels of wino-type neutralino and in the lower panel, we have shown wino-type chargino decay channels.
The upper and lower panel correspond to the decay channels of wino-type $\lspone$ and $\chonepm$ respectively.

Also, we have shown the variation of branching ratios with $\mlspone$ and $\mchonepm$ in Figure~\ref{fig:br_mass}. 
%According to the Eqs.~(4.3)-(4.5) of Ref.~\cite{Dumitru:2018nct} and Eqs.~(70)-(74) in Ref.~\cite{FileviezPerez:2012mj}, the branching ratios corresponding $Z$ and $W$ boson will increase with increasing mass of chargino and neutralino. 
We observe that the branching ratios corresponding to $Z$ and $W$ boson increase with increasing $\mlspone/\mchonepm$ as reported in previous literature~\cite{FileviezPerez:2012mj,Dumitru:2018nct}.
%%%%%%%%%%%%%%%%%%%%%%%%%%%%
\begin{figure}[!htb]
\includegraphics[width=1\textwidth]{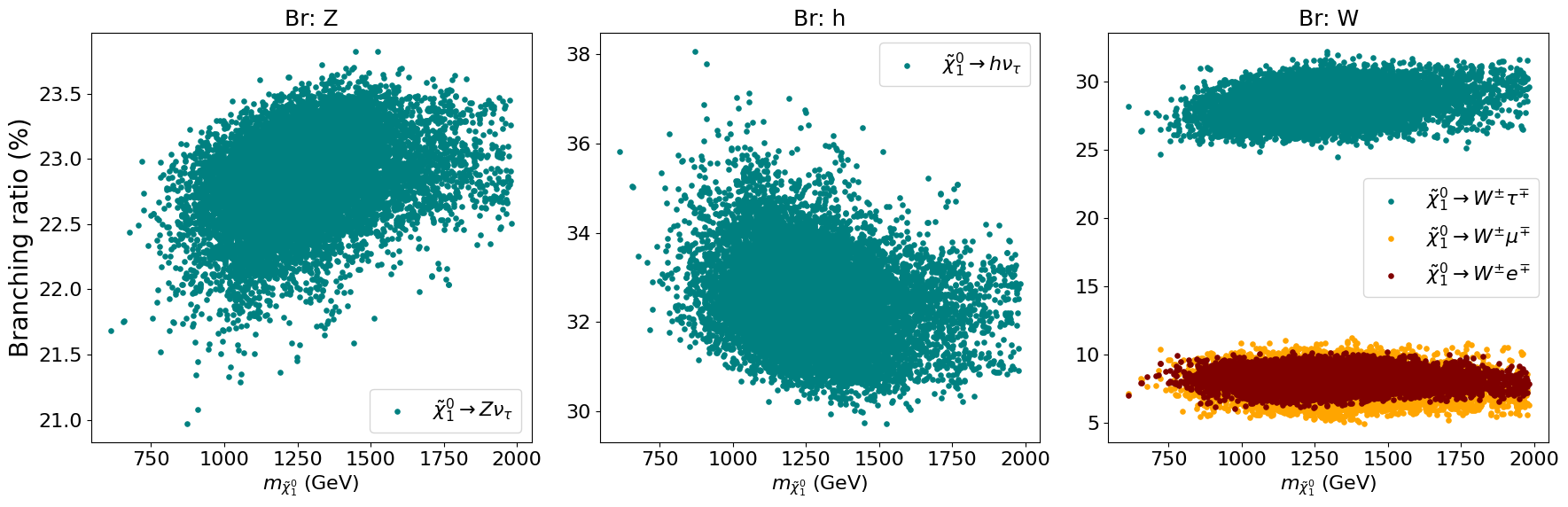}
\includegraphics[width=1\textwidth]{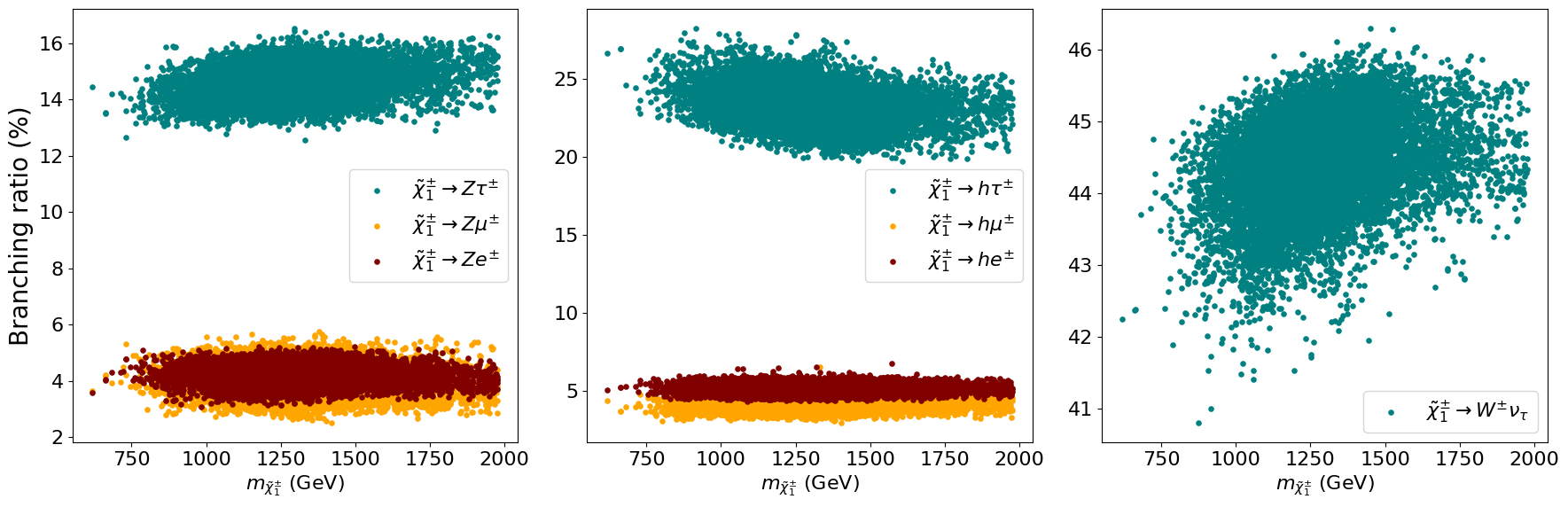}
    \caption{
    The branching ratios of $\lspone$ and $\chonepm$ into different lepton flavors are shown. In the upper panel, $\mathcal{B}r(\lspone \rightarrow Z\nu)$ (left), $\mathcal{B}r(\lspone \rightarrow h\nu)$ (middle), and $\mathcal{B}r(\lspone \rightarrow W^{\pm}l^{\mp})$ (right) are presented. In the lower panel, $\mathcal{B}r(\chonepm \rightarrow Zl^{\pm})$ (left), $\mathcal{B}r(\chonepm \rightarrow hl^{\pm})$ (middle), and $\mathcal{B}r(\chonepm \rightarrow W^{\pm}\nu)$ (right) are shown. The red, yellow, and teal colors correspond to the first-, second-, and third-generation lepton flavors, respectively.
    %Upper Panel: The branching ratios to different lepton flavors are shown for the neutralino LSP $\lspone$, corresponding to the decay modes: $\mathcal{B}r(\lspone \rightarrow Z\nu)$ (left), $\mathcal{B}r(\lspone \rightarrow h\nu)$ (middle), and $\mathcal{B}r(\lspone \rightarrow W^{\pm}l^{\mp})$ (right). Lower Panel: The branching ratios to different lepton flavors are shown for the chargino LSP $\chonepm$, corresponding to the decay modes: $\mathcal{B}r(\chonepm \rightarrow Zl^{\pm})$ (left), $\mathcal{B}r(\chonepm \rightarrow hl^{\pm})$ (middle), and $\mathcal{B}r(\chonepm \rightarrow W^{\pm}\nu)$ (right). 
   }
    \label{fig:br_flavor}
\end{figure}
%%%%%%%%%%%%%%%%%%%%%%%%%%%%%%%%%%%%%%%%%
We now illustrate the effect of the neutrino mass hierarchy on different neutrino and lepton flavors corresponding to various decay channels of the neutralino and chargino in Figure~\ref{fig:br_flavor}. The upper and lower panels represent the decay modes of the lightest neutralino and lighter chargino, respectively. %\tcr{Here also either mention $\lspone$ or the lightest neutralino , and/or  the lighter chargino or symbol.} 
The teal, yellow, and red colors refer to the $\tau$, $\mu$, and $e$-flavored neutrino and lepton in Figure~\ref{fig:br_flavor}.
Since the third neutrino, predominantly $\tau$-flavored, is the heaviest, the \texttt{bRPV} coupling $\epsilon_3$ has the largest value and the decay channels $\lspone \rightarrow Z\nu_{\tau}$, $\lspone \rightarrow h\nu_{\tau}$ dominate. Similarly, for the chargino, the $\chonepm \rightarrow W^{\pm} \nu_{\tau}$ decay channel prevails over those involving other neutrino flavors.
Furthermore, due to the hierarchy $m_{\nu_3} > m_{\nu_2} > m_{\nu_1}$, the $\lspone \rightarrow W^{\pm}\tau^{\mp}$ decay channel is significantly more prominent than $\lspone \rightarrow W^{\pm}\mu^{\mp}$ and $\lspone \rightarrow W^{\pm}e^{\mp}$, as depicted in the right-hand plot of the upper panel in Figure~\ref{fig:br_flavor}. A similar pattern is observed for the $\chonepm \rightarrow Z\tau^{\pm}$ and $\chonepm \rightarrow h\tau^{\pm}$ channels, as seen in the left and middle plots of the lower panel in Figure~\ref{fig:br_flavor}, respectively.

\subsubsection{Revisiting the LHC trilieton resonance search}
\label{sec:atlas_search}

\begin{figure}[!htb]
\includegraphics[width=0.49\textwidth]{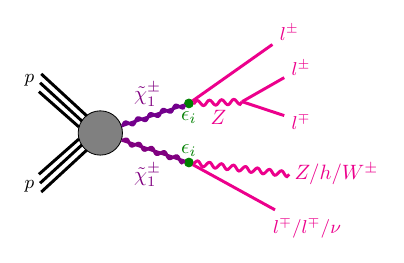}
\includegraphics[width=0.49\textwidth]{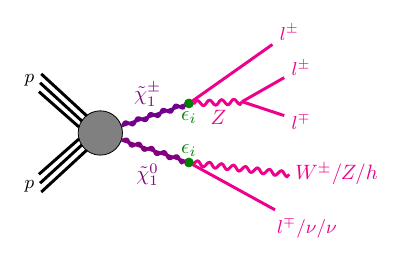}
    \caption{Diagrams of $\chonepm\chonemp$ (left) and $\chonepm\lspone$ (right) productions and decay where atleast one $\chonepm$ decays like $\chonepm \to Zl \to lll$. The other $\chonepm$ and $\lspone$ decay to $W$, $Z$ and $h$ boson according to bilinear RPV coupling $\epsilon$. 
    %The figures are taken from the Ref.~\cite{ATLAS:2020uer}. \tcm{instead can you draw the FD ?}
    }
    \label{fig:decay_atlas}
\end{figure}

ATLAS Collaboration has searched for wino-type $\chonepm\chonemp$ + $\chonepm\lspone$ pair production at $\sqrt{s} = 13$ TeV with $\mathcal{L} =$ 139 fb$^{-1}$~\cite{ATLAS:2020uer}. They have looked for trilepton invariant mass spectrum resonance ($m_{Zl}$), which comes from the wino pair productions where at least one $\chonepm$ decays to $Zl^{\pm}$. The limit on $m(\chonepm) = m(\lspone)$\footnote{In our case, the masses are not exactly equal but are nearly degenerate and both the particles decay via RPV SUSY interactions.} %\tcm{( .... Here add a footnote that in reality/our case the masses are not exactly equals. They are nearly mass degenerate... both decay via RPV couplings)} 
was presented as a function of their decay branching ratio to the $Z$ boson. The decay channels considered by ATLAS corresponding to $\chonepm\chonemp$ and $\chonepm\lspone$ are shown in Figure~\ref{fig:decay_atlas}. One reconstructed $\chonepm$ is required in the signal event. Also, events are categorized into three signal regions (\texttt{SR}s: \texttt{SR3l}, \texttt{SR4l}, and \texttt{SRFR}) depending on the number of leptons and the presence of a second reconstructed $W$, $Z$, or $h$ boson from second $\chonepm$/$\lspone$ decay. The \texttt{SRFR} region targets events where all decay products are visible and ``fully reconstructed''. The \texttt{SR4l} region targets events with four or more leptons and possible missing energy ($\met$), while the \texttt{SR3l} region targets events with only three visible leptons and substantial $\met$, with at least one neutrino coming from the decay of the second $\chonepm$/$\lspone$.

%%%%%%%%%%%%%%%%%%%%%%%%%%%%

%%%%%%%%%%%%%%%%%%%%%%%%%%%%%%%%%%%%%%%%%
The ATLAS collaboration has observed that the three signal regions contribute roughly equally to the overall sensitivity of the search, with a minor increase in sensitivity to Higgs boson decays from \texttt{SRFR} offset by a similar increase in sensitivity to $W$ boson decays from \texttt{SR4l} (see Sec~8.2 of the Ref.~\cite{ATLAS:2020uer}). 
%%%%%%%%%%%%%%%%%%%%%%%%%%%%%%%%%%%%%%%%
\begin{table}[!htb]
\centering
\scriptsize
\renewcommand{\arraystretch}{1.3} % Adjust the row spacing
\resizebox{\textwidth}{!}{
\begin{tabular}{|c|c|c|c|c|c|c|c|c|} 
    \hline
    Mass & \multicolumn{4}{c|}{500 GeV} & \multicolumn{4}{c|}{800 GeV}  \\
    \hline
     \multirow{2}{*}{Selection} & \multicolumn{2}{c|}{$\chonepm\lspone$} & \multicolumn{2}{c|}{$\chonepm\chonemp$} & \multicolumn{2}{c|}{$\chonepm\lspone$} & \multicolumn{2}{c|}{$\chonepm\chonemp$}\\
    \cline{2-9}
    & ATLAS & Our & ATLAS & Our & ATLAS & Our & ATLAS & Our \\
    \hline
    Total production ($\sigma \times \mathcal{L}$) & 6440 & 6442 & 3070 & 3074 & 661 & 661 & 307 & 307\\
    \hline
%    Generation filters ($Z/3l$) & 171 & -& 214 & - & 18.5 & - & 21.0 & -\\
%    \hline
%    Preliminary event reduction & 169 & - & 189 & - & 18.3 & - & 19.0 & -\\
%    \hline
%    Triggering & 168 & - & 187 & - & 18.1 & - & 18.9 & -\\
%    \hline
    $\geq 3$ signal leptons & 82.1 & 96.56 & 74.4 & 96.23 & 8.89 & 11.21 & 7.66 & 10.65\\
    \hline
    $Z$ candidate & 63.0 & 46.81 & 52.0& 39.29 & 6.79 & 5.44 & 5.41 & 5.00\\
    \hline
    SR$3l$ assignment & 49.5 & 38.48 &30.8 & 26.69 & 5.29 & 4.38 & 3.16 & 3.22\\
    \hline
    $\met > 150$ GeV & 33.4 & 26.32 &15.2 & 11.93 & 4.41 & 3.46 & 2.03 & 1.92\\
    \hline
    $m_T^{\text{min}} > 125$ GeV & 25.8 & 22.49 & 11.2& 9.03 & 3.65 & 2.98 & 1.59& 1.60\\
    \hline
    $\Delta R(b_1,b_2) < 1.5$ & 24.8 & 21.26 & 10.7& 8.79 & 3.54 & 2.91 & 1.49 & 1.58\\
    \hline
%    MC-to-data eff. weights & 23.5 & - & 10.0& -& 3.34 & -& 1.41& -\\
%    \hline
    \texttt{SR3l}$_{e\mu}$ & 23.4 & 21.26 & 10.04& 8.79& 3.34 & 2.91& 1.41& 1.579\\
    \hline
    \texttt{SR3l}$_e$ & 11.9 & 9.95 & 4.85 & 4.00 & 1.71 & 1.16& 0.711& 0.676\\
    \hline
    \texttt{SR3l}$_{\mu}$ & 11.5 & 11.31 & 5.19& 4.79 & 1.63 & 1.75 & 0.703& 0.903\\
    \hline
\end{tabular}
}
\caption{The signal yields provided by both ATLAS and derived from our simulation corresponding to \texttt{SR3l} signal region at $\mchonepm =\mlspone =$~500 GeV \& 800 GeV. %\tcm{should we merge the last two rows ?}
} 
\label{tab:atlas_validation}
\end{table}
In this work, we  focus only on the \texttt{SR3l} signal region. First, we validate the result quoted in Table 5 of the auxiliary material provided by the ATLAS collaboration \cite{atlas_wino_web}, where democratic branching fractions into bosons ($W$, $Z$, and Higgs) and leptons ($e$, $\mu$, and $\tau$) are considered. We generate our signal events using Monte Carlo event generation with \texttt{MadGraph}~\cite{Alwall:2014hca} and do detector simulation with \texttt{Delphes-3.5.0}~\cite{deFavereau:2013fsa}. We use the cross-sections at 13 TeV provided by the LHC SUSY working group~\cite{Fuks:2012qx,Fuks:2013vua}. Using this set-up, we have validated our results for two mass points, such as 500 GeV and 800 GeV, and the comparison of the two results is shown in the Table~\ref{tab:atlas_validation}. We observe an overall good agreement between our simulation and the ATLAS results for the signal yields.

\begin{figure}[!htb]
\centering
\includegraphics[width=0.6\textwidth]{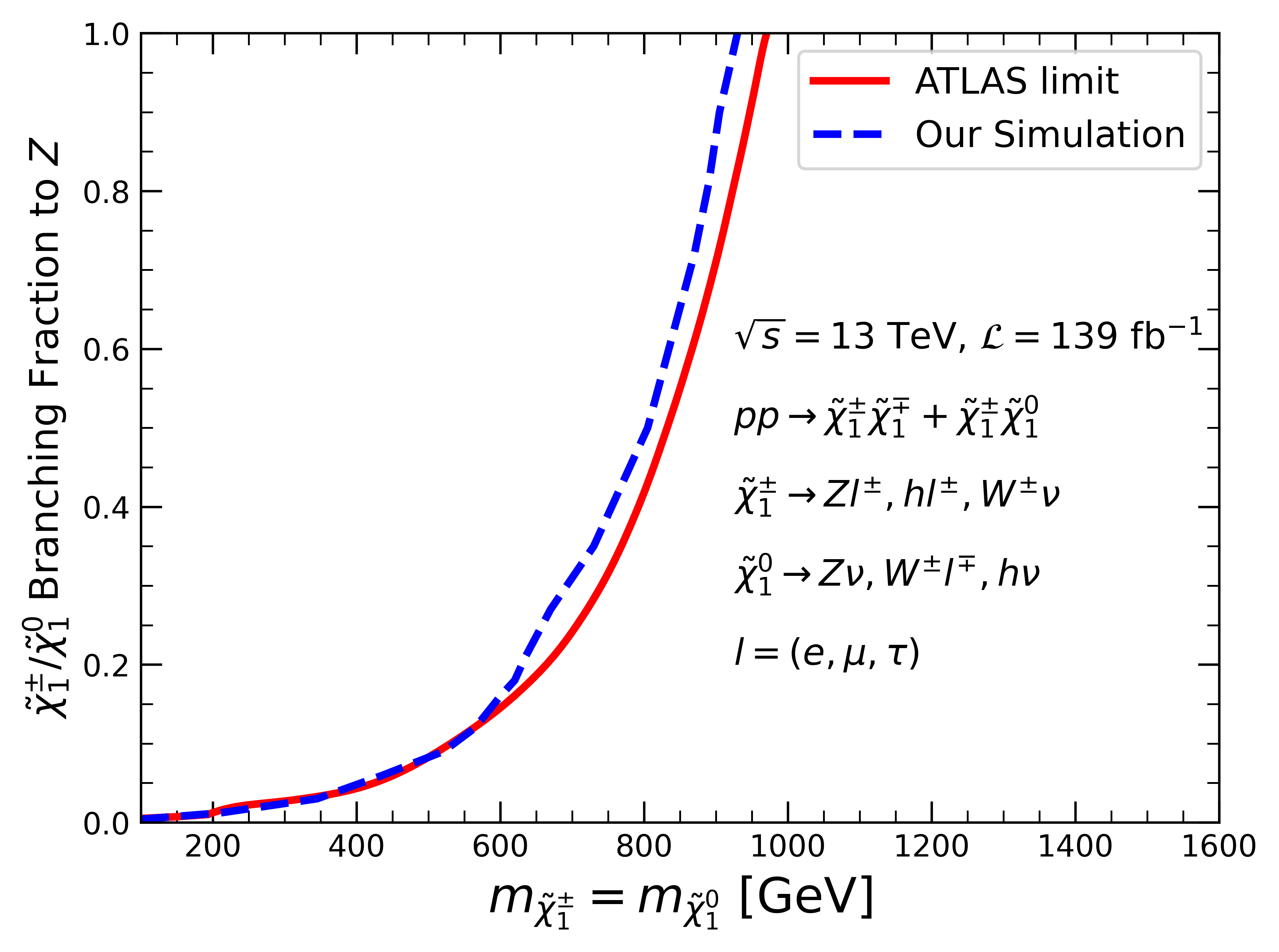}
    \caption{Comparison of the 95\% C.L. exclusion limit obtained by ATLAS collaboration with Run-II data~\cite{ATLAS:2020uer} and our analysis. The red colored solid line represents the limit obtained by ATLAS and the blue dashed line corresponds to the exclusion reach obtained from our analysis. Here $l=e, \mu, \tau$ is considered and the charged-lepton decays of $\chonepm$/$\lspone$ into any leptons with equal probability is imposed here. The sum of the $\chonepm$/$\lspone$ branching fractions to $W$, $Z$, and $h$ bosons is unity for each point, and the branching fractions to $W$ and $h$
bosons are chosen so as to be equal everywhere.}
    \label{fig:br_Z_limit}
\end{figure}
%%%%%%%%%%%%%%%%%%%%%%%%%%%%%%%%%%%%%%%%%

The 95\% CL upper limits on number of signal events ($S^{95}_{\text{obs}}$) for each $m_{Zl}$ mass bin of every SR is provided by the ATLAS Collaboration (see Table~4 of Ref~\cite{ATLAS:2020uer}). As mentioned above, we consider only \texttt{SR3l} signal region and depending on these given upper limits, we validate the ATLAS exclusion limit 
on the $\chonepm$/$\lspone$ decay branching fraction to $Z$ and $\chonepm/\lspone$ mass plane.
In Figure~\ref{fig:br_Z_limit}, we present our validation result where the red solid line represents the ATLAS exclusion limit and the dashed blue line refers to the exclusion line generated by our simulation. Here, all three flavors of leptons ($e, \mu, \tau$) have equal branching ratios. The sum of the $\chonepm$/$\lspone$ branching fractions to $W$, $Z$, and $h$ bosons is unity for each point, and the branching fractions to $W$ and $h$ bosons are considered to be equal everywhere. From Figure~\ref{fig:br_Z_limit}, it is evident that our result overlaps at the lower mass region (below $\sim$ 600 GeV) and we also have a nice agreement with the experimental result at higher mass region. The slight difference at the higher mass region may arise because the ATLAS exclusion curve was derived using three signal regions, whereas our result is based on \texttt{SR3l} signal region.

All the above results are for simplified scenarios with some assumptions. In contrast, when we satisfy neutrino oscillation data, the branching ratios to different flavors are generally unequal as discussed in Section~\ref{sec:nh}. The masses and the branching ratios of $\lspone$ and $\chonepm$ at the best-fit point are shown in the Table~\ref{tab:best_fit}. It is clear from this table that we have a slight ($\sim$3 GeV) mass difference between $\mchonepm$ and $\mlspone$, and the branching ratios to $Z$ boson are not equal for $\chonepm$ and $\lspone$ because of the neutrino oscillation data. At the best-fit point, we have obtained that we can exclude $\chonepm$ upto $\sim$565 GeV with 23\% branching ratio to $\chonepm \to Ze + Z\mu + Z\tau$ decay.

%In this plane we also show the points within $3\sigma$ allowed parameter space coming from the neutrino data where all these data points lie in the allowed region. The black dashed line at around 575 GeV is the exclusion when we consider the branching ratio of the best-fit point as shown in Table.~\ref{tab:best_fit}.

\subsubsection{Prospects of Trilepton Resonance Search at the HL‑LHC}
\label{sec:hllhc}

%%%%%%%%%%%%%%%%%%%%%%%%%%%
We now show the sensitivity reach at the HL‑LHC ($\sqrt{s} = 14$~TeV, $\mathcal{L} = 3$~ab$^{-1}$) for wino-type $\chonepm\chonemp$ + $\chonepm\lspone$ pair production. We apply the cuts used for \texttt{SR3L} signal region from ATLAS paper and also assume that the signal selection efficiencies remain unchanged for backgrounds. The predicted exclusion reaches at the HL‑LHC are obtained by scaling the background yields from the current ATLAS analysis~\cite{ATLAS:2020uer} to the higher luminosity and center-of-mass energy. We assume that the cross sections for the background at 14 TeV will be increased by a factor of 1.2. As similar to 13 TeV results discussed in Section~\ref{sec:atlas_search}, the signal events are generated using \texttt{MadGraph}~\cite{Alwall:2014hca} and fast detector simulation is done with \texttt{Delphes-3.5.0}~\cite{deFavereau:2013fsa}. The NLO+NLL signal production cross-sections at 14 TeV are estimated by Resummino~\cite{Fuks:2012qx,Fuks:2013vua}.
%%%%%%%%%%%%%%%%%%%%%%%%%%%%
\begin{figure}[!htb]
\centering
\includegraphics[width=0.6\textwidth]{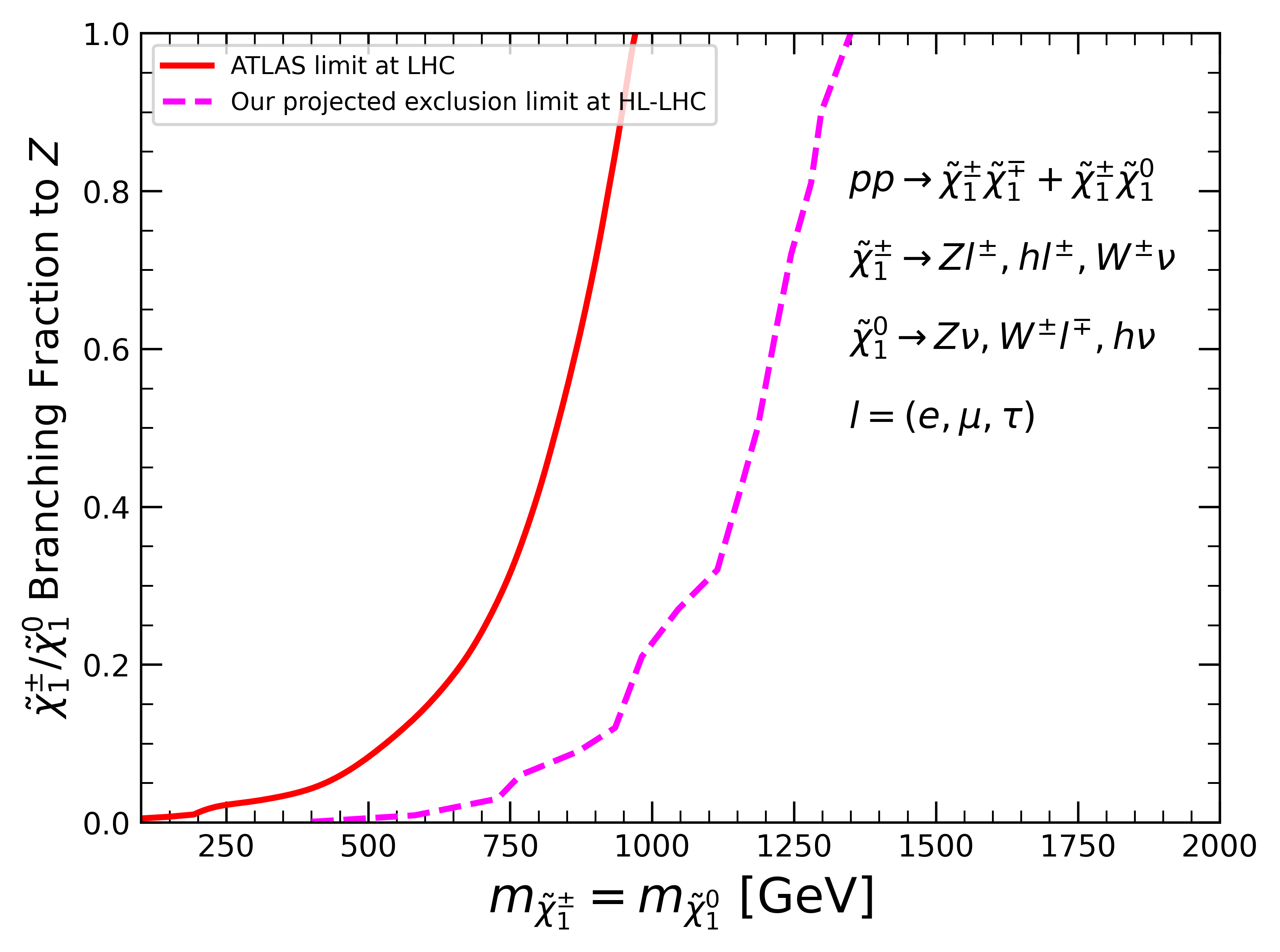}
    \caption{The 95\% C.L. projected exclusion limit obtained by our simulation at HL-LHC is shown by magenta-colored dashed line. The red colored solid line represents the limit obtained by ATLAS with Run-II data. Here $l=e, \mu, \tau$ is considered, and the charged-lepton decays of $\chonepm$/$\lspone$ into any leptons with equal probability are imposed here. The sum of the $\chonepm$/$\lspone$ branching fractions to $W$, $Z$, and $h$ bosons is unity for each point, and the branching fractions to $W$ and $h$
bosons are considered to be equal everywhere.}
    \label{fig:Z_limit_14}
\end{figure}
%%%%%%%%%%%%%%%%%%%%%%%%%%%%%%%%%%%%%%%%%
%to get the signal cross-sections at 14 TeV, we use Resummino~\cite{Fuks:2012qx,Fuks:2013vua} \tcb{NNLO+.. signal cross-sections at 14 TeV are estimated by Resummino~\cite{Fuks:2012qx,Fuks:2013vua}}. 
%\tcm {!!!!! for signal or BKG?}. 
We find the exclusion mass value by applying the condition 
\begin{equation}
\frac{S}{\sqrt{B + (\Delta B\times B)^2}} > 2,
\end{equation}
where $S$, $B$, and $\Delta B$ are the calculated signal yield, the scaled background yield, and the uncertainty in the background yield, respectively. We consider 10\% uncertainty in the background yield here and show the projected exclusion limit at HL-LHC obtained by our simulation in Figure~\ref{fig:Z_limit_14}. Here, the magenta dashed line represents the HL-LHC projected exclusion limit and the red solid line refers to the ATLAS limit with Run-II data. For branching ratios of $\chonepm$ decays into a $Z$ boson  and a lepton of approximately 1\%, 50\%, and 100\%, wino masses can be excluded up to about $600~\mathrm{GeV}$, $1185~\mathrm{GeV}$, and $1350~\mathrm{GeV}$, respectively.
%With branching ratio to $Z$ with value $\sim$ 0.1\%, 50\%, and 100\%, we can exclude $\mchonepm = \mlspone$ upto around 400 GeV, 1185 GeV, and 1350 GeV respectively. 
%\tcr{At the best-fit point, with 23\% branching ratio of $\chonepm \to Ze + Z\mu + Z\tau$ decay, one can exclude $\chonepm$ upto $\sim$950 GeV at HL-LHC. It is clear from the distribution of $M_2$ shown in Figure~\ref{fig:param_region} that we can probe only part of the allowed parameter space at HL-LHC.}
Considering the branching ratio
$\mathrm{Br}(\chonepm \to Zl^\pm; l= e,\mu,\tau) \sim$23\%, obtained at the best-fit point, 
the wino-like mass degenerate $\lspone/\chonepm$ can be excluded upto 950 GeV at the HL-LHC.
Our analysis shows that the HL-LHC can probe a significant portion of 1$\sigma$ allowed parameter space (see the distribution of $M_2$ shown in Figure~\ref{fig:param_region})  by neutrino oscillation measurements and other experimental constraints.

\subsection{Inverted Hierarchy: Brief Discussion}
\label{sec:ih}
%%%%%%%%%%%%%%%%%%%%%%%%%%%
\begin{figure}[!htb]
\includegraphics[width=1\textwidth]{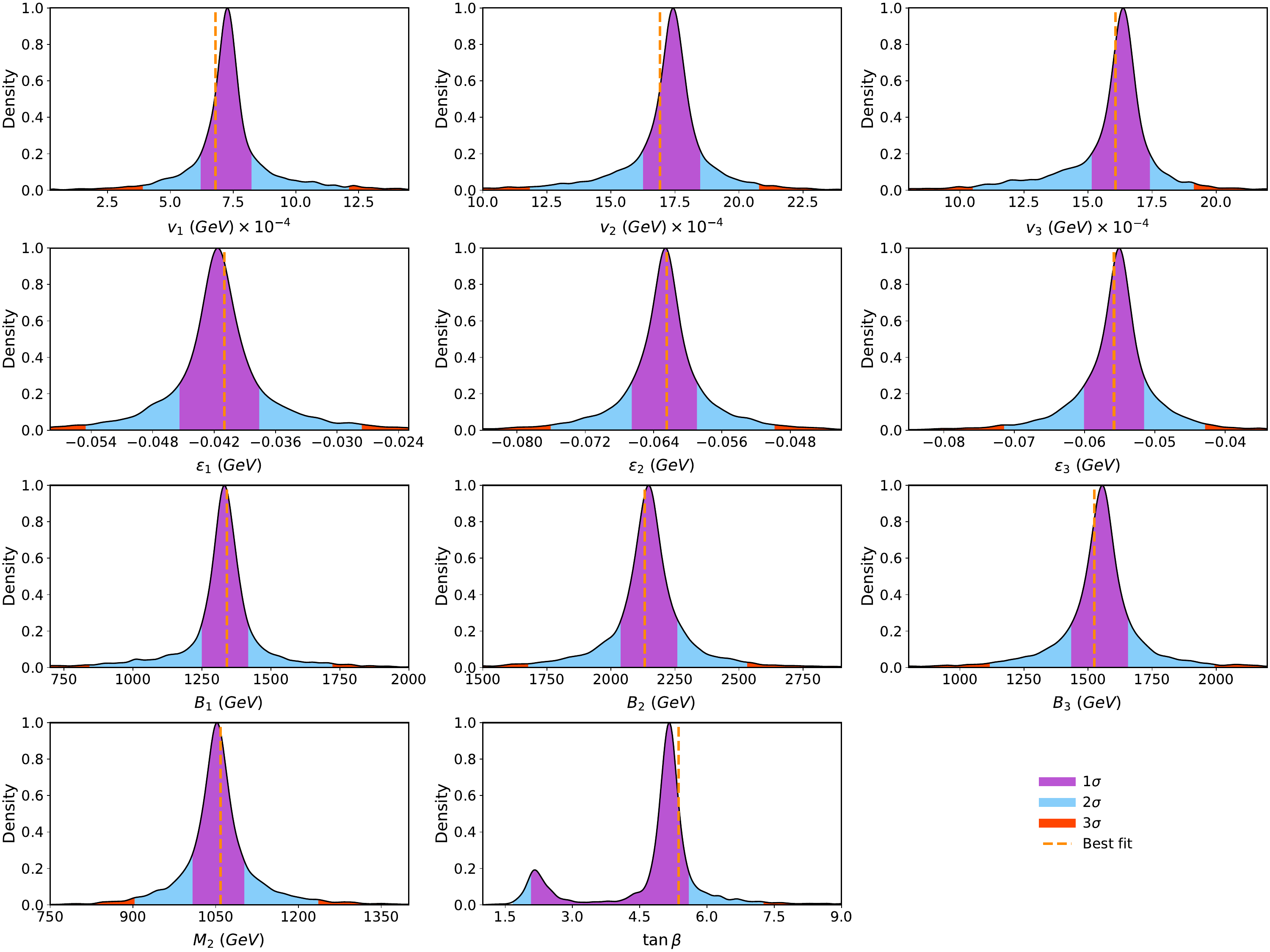}
    \caption{The 1-D posterior distribution of all the input parameters for IH scenario along with $1\sigma$, $2\sigma$ and $3\sigma$ regions are shown here with purple, cyan and red colors respectively. The best-fit value of each parameter is represented by the dashed orange colored line.}
    \label{fig:param_region_ih}
\end{figure}
%%%%%%%%%%%%%%%%%%%%%%%%%
%%%%%%%%%%%%%%%%%%%%%%%%%%%%%%%%%%%%%%%%%
\begin{table}[!htb]
\centering
\resizebox{\textwidth}{!}{
\begin{tabular}{|c|c|c|c||c|c|c|c|} 
    \hline
    \multicolumn{8}{|c|}{Best-fit Point (BFP)}\\
    \hline
    \multicolumn{4}{|c||}{Input parameters} &  \multicolumn{2}{c|}{$m_{\lspone}$ = 1102 GeV} & \multicolumn{2}{c|}{$m_{\chonepm}$ = 1106 GeV} \\
    \hline
    Parameter & Value & Parameter & Value & Decay & Br (\%) & Decay & Br(\%) \\
	\hline
	$M_2$ [GeV] & 1058.49 & $\epsilon_2$ [$10^{-2}$GeV] & -6.24 & $We$ & 0.16 & $W\nu_{\mu}$ & 43.99 \\
	%\hline
	$\tan\beta$ & 5.36 & $\epsilon_3$ [$10^{-2}$GeV] & -5.58  & $W\mu$& 21.46& $he$ & 0.74 \\
   % \hline
    $v_1$ [$10^{-4}$GeV] & 6.79 & $B_1$ [GeV] & 1341  & $W\tau$& 22.36 & $h\mu$& 17.12 \\
    %\hline
    $v_2$ [$10^{-3}$GeV] & 1.69 & $B_2$ [GeV] & 2131 & $h\nu_e$ & 0.28 & $h\tau$ & 15.57 \\
    %\hline
    $v_3$ [$10^{-3}$GeV] & 1.60 & $B_3$ [GeV] & 1524   & $h\nu_{\mu}$& 33.10 & $Ze$ & 0.08 \\
    $\epsilon_1$ [$10^{-2}$GeV] & -4.09 &  &  & $h\nu_{\tau}$& 0.02 & $Z\mu$& 11.02 \\
    \cline{1-4}
    \multicolumn{4}{|c||}{$\chi^2_{\text{min}}/\texttt{d.o.f} = 3.94/4 = 0.98$} & $Z\nu_{\mu}$ & 22.62 & $Z\tau$& 11.48 \\
    \hline
\end{tabular}
}
\caption{The input parameters at the best-fit point, along with the masses of the wino-type lightest neutralino and chargino and their respective decay branching ratios, are presented corresponding to IH scenario.
}
\label{tab:best_fit_ih}
\end{table}
%%%%%%%%%%%%%%%%%%%%%%%%%%%%
\begin{figure}[!htb]
\includegraphics[width=1\textwidth]{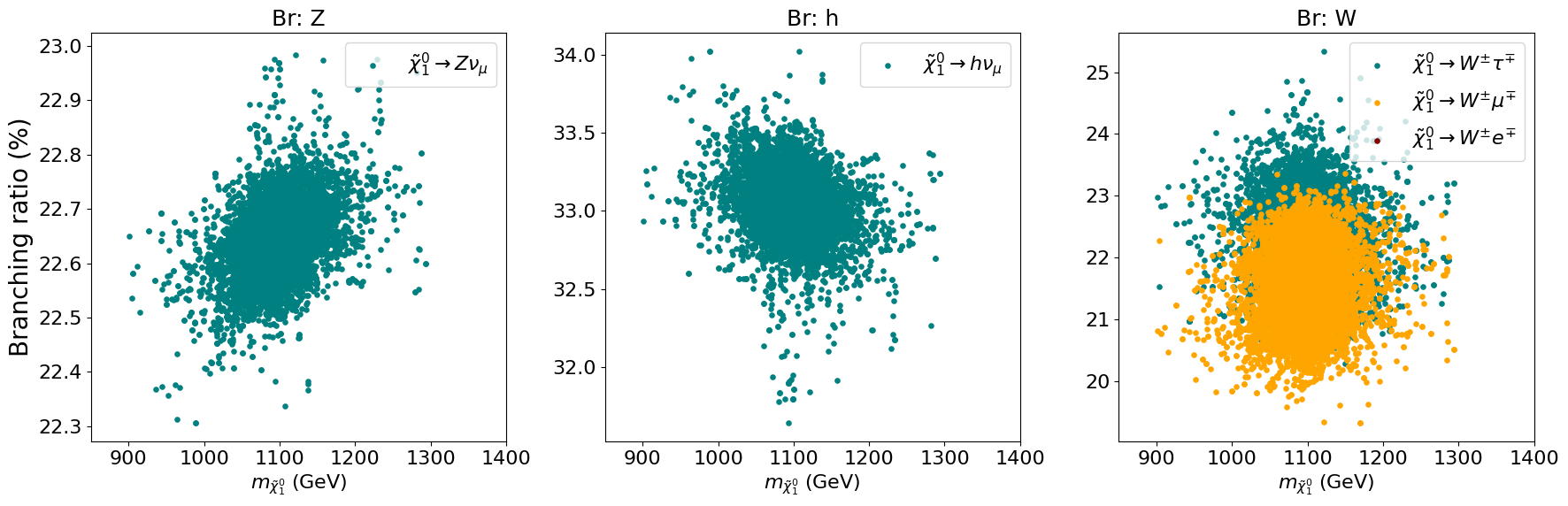}
\includegraphics[width=1\textwidth]{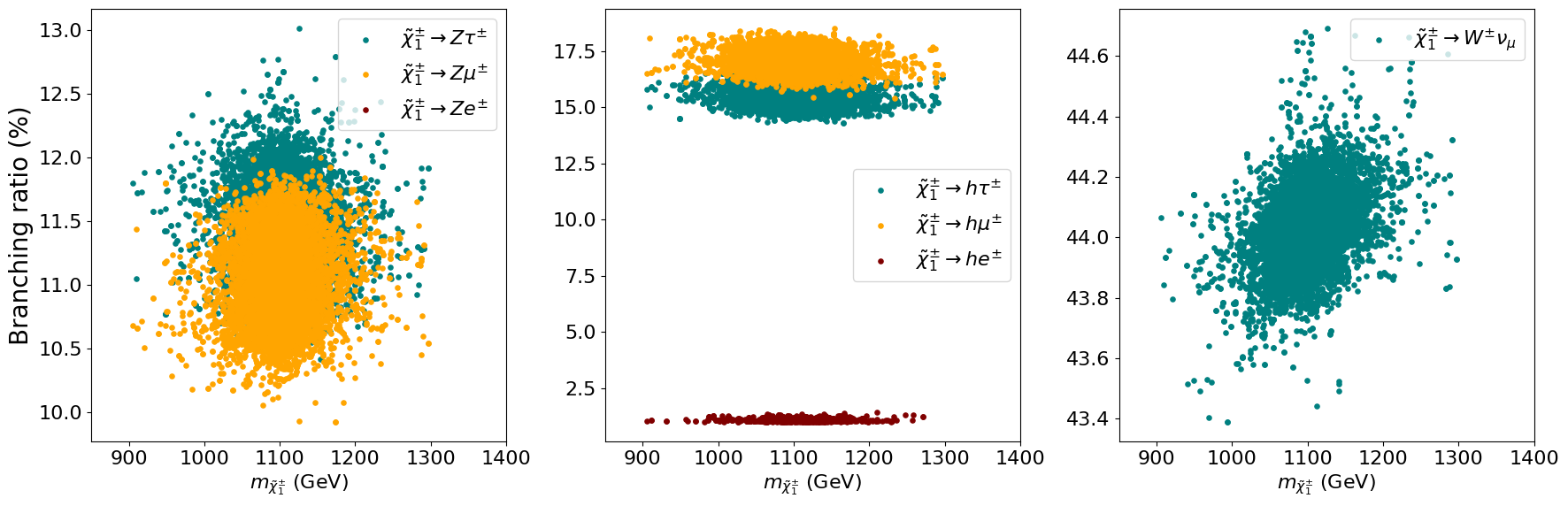}
    \caption{The branching ratios of $\lspone$ and $\chonepm$ into different lepton flavors are shown. In the upper panel, $\mathcal{B}r(\lspone \rightarrow Z\nu)$ (left), $\mathcal{B}r(\lspone \rightarrow h\nu)$ (middle), and $\mathcal{B}r(\lspone \rightarrow W^{\pm}l^{\mp})$ (right) are presented. In the lower panel, $\mathcal{B}r(\chonepm \rightarrow Zl^{\pm})$ (left), $\mathcal{B}r(\chonepm \rightarrow hl^{\pm})$ (middle), and $\mathcal{B}r(\chonepm \rightarrow W^{\pm}\nu)$ (right) are shown. The red, yellow, and teal colors correspond to the first-, second-, and third-generation lepton flavors, respectively.}
    \label{fig:br_flavor_ih}
\end{figure}
%%%%%%%%%%%%%%%%%%%%%%%%%%%%%%%%%%%%%%%%%
We perform a similar analysis for the IH case where the second neutrino mass eigenstate is the heaviest one and the third neutrino mass eigenstate is the lightest one. We follow the same procedure for IH also as considered for the NH scenario and we present the allowed parameter space in Figure~\ref{fig:param_region_ih}. 
The allowed region of parameter space is significantly smaller compared to the NH scenario, which arises due to the structure of the neutrino mass hierarchy. This hierarchy is also reflected in the best-fit and allowed values of the model parameters as shown in Table~\ref{tab:best_fit_ih}. As mentioned above, the second neutrino state is the heaviest one and it has an almost equal admixture of all the three neutrino flavors. So, the contribution from tree-level corresponding to any single flavor must not be very large. This leads to the smaller $\tan\beta$ and $M_2$ as compared to the NH scenario as shown in Table~\ref{tab:best_fit_ih} as well as the allowed regions in Figure~\ref{fig:param_region_ih}. The $3\sigma$ allowed regions corresponding to $\tan\beta$ and $M_2$ are 1.5-8.0 and 850-1300 GeV respectively. 
%\tcr{ A detailed discussion of the hierarchy of all the input parameters and the comparison with NH case can be found in our previous work~\cite{Choudhury:2023lbp}. .. may be removed}

The hierarchy of neutrino mass affects the \texttt{bRPV} parameters and controls the pattern of branching ratios of the chargino and neutralino decays into different lepton flavors, as shown in Table~\ref{tab:best_fit_ih}. The decay branching ratios into different bosons accompanied by a lepton or neutrino are illustrated in Fig.~\ref{fig:br_flavor_ih}. In the IH scenario, the values of the \texttt{bRPV} parameters favor the muon flavor, and therefore the dominant decay channels are $\lspone \to Z \nu_{\mu}$ and $\lspone \to h \nu_{\mu}$, whereas the other two flavors vanish as discussed in the Ref.~\cite{Dumitru:2018nct}. Similarly, for the chargino decay $\chonepm \to W^{\pm} \nu_{\mu}$ becomes the dominant mode.

From Tables~\ref{tab:best_fit} and \ref{tab:best_fit_ih}, we observe that the total branching ratio $\mathrm{Br}(\chonepm \to Zl)$ ($l=e,\mu,\tau$) is approximately 23\% for both the NH and IH scenarios. However, the flavor composition of this decay differs significantly between the two cases. For NH, the decay $\chonepm \to Z\tau$ dominates with a branching ratio of about 14\%, while $\chonepm \to Z\mu$ and $\chonepm \to Ze$ each have branching ratios of roughly 4\%. In contrast, for the IH, the branching ratios of $\chonepm \to Z\tau$ and $\chonepm \to Z\mu$ are both around 11\%, whereas $\chonepm \to Ze$ is highly suppressed. Therefore, although the inclusive $\chonepm \to Zl$ rate is similar in the two scenarios, the number of events in flavor-specific channels can differ substantially. In particular, the $\mu$ channel is expected to yield significantly more signal events in the IH case than in the NH case. Consequently, flavor-tagged collider searches may provide a means to distinguish between the NH and IH neutrino mass hierarchies.

%\tcm{you may add few lines about the collider prospect.. whether the reach will change or not.. what mode will give more signal yield etc,,,}

%%%%%%%%%%%%%%%%%%%%%%%%%%%%%%%%%%%%%%%%%

\section{Conclusion}
\label{sec:conclusion}
%%%%%%%%%%%%%%%%%%%%%%%%%%%%%%%%
Neutrino oscillation experiments have conclusively demonstrated that the three light neutrinos possess non-zero masses and mix with each other, a feature that the Standard Model (SM) cannot explain. Supersymmetry (SUSY) provides a well-motivated framework for new physics, and in the bilinear R-parity violating (\texttt{bRPV}) SUSY scenario, neutrino masses can be generated naturally without introducing additional particles. In this work, we have investigated  neutrino mass generation in the context of \texttt{bRPV} SUSY model. By combining neutrino oscillation data, the observed Higgs mass and its coupling strength modifiers, and flavor physics observables such as $B$-hadron decay branching ratios, we have conducted an MCMC scan to identify the allowed parameter space for this model parameters while satisfying the current LHC limits. From this allowed region, we selected representative points to illustrate the decay branching ratios of wino-like lighter charginos ($\chonepm$) and lightest neutralinos ($\lspone$). We find that the branching ratios to different neutrino and charged lepton flavors depend sensitively on the neutrino mass hierarchy. 
%\tcm{and the theoretical predictions agree well with our computed results.!!!!} 
For the Normal Hierarchy scenario, the decay channel producing $\tau$-flavored neutrino together with a SM boson is the dominant mode, whereas in the Inverted Hierarchy case, the decay producing $\mu$-flavored neutrino becomes the dominant one. 

Furthermore, we explore the current LHC bounds and future search sensitivity from 
the trilepton resonance searches on the allowed parameter space. We validate the exclusion limits provided by the ATLAS Collaboration for wino-like $\lspone$ or $\chonepm$ using trilepton resonance search with Run-II data. We showed that our simulated result has a nice agreement with the experimental limit. 
Using the branching ratio at the best-fit point $\mathrm{Br}(\chonepm \to Zl^\pm) \sim$23\%, coming from our analysis with neutrino oscillation data along with Higgs data and flavor data, we obtain the current LHC exclusion on the wino-like chargino is around 565 GeV from LHC Run-II data. We also estimate the projected exclusion reach at HL-LHC with $\sqrt{s}=$ 14 TeV and $\mathcal{L} =$ 3 ab$^{-1}$. The projected exclusion reach with a similar branching ratio
at High-Luminosity LHC (HL-LHC) is around 950 GeV. For the decay of $\chonepm$/$\lspone$ into a $Z$ boson and a lepton with branching ratios of 1\%, 50\%, and 100\%, wino masses can be excluded up to $\sim$ $600~\mathrm{GeV}$, $1185~\mathrm{GeV}$, and $1350~\mathrm{GeV}$ respectively.
%For branching ratios of $\chonepm$/$\lspone$ decays into a $Z$ boson and a lepton of 1\%, 50\%, and 100\%, wino masses can be excluded up to about $600~\mathrm{GeV}$, $1185~\mathrm{GeV}$, and $1350~\mathrm{GeV}$, respectively.
Our analysis shows that the HL-LHC can probe a significant portion of the 1$\sigma$ allowed parameter space  by neutrino oscillation measurements and other experimental constraints.
Overall, our results demonstrate the interplay between neutrino physics and collider observables in constraining \texttt{bRPV} SUSY and provide a road-map for testing these scenarios at upcoming experiments.

\section{Acknowledgement}
\label{sec:ack}
%%%%%%%%%%%%%%%%%%%%%%%%%%%%%%%%%
%AC and AM acknowledge ANRF India for Core Research Grant no. CRG/2023/008570. The authors also  thank Subhadeep Mondal and Sourav Mitra for their insightful discussions and contributions  regarding  the analysis setup. AM acknowledges Sabine Kraml for her careful reading of the manuscript and for providing constructive feedback. The work of AM is supported by the French Agence Nationale de la Recherche under grant ANR-23-CHRO-0006 (OpenMAPP).

AC and AM acknowledge support from the ANRF India through the Core Research Grant No. CRG/2023/008570. The authors are grateful to Subhadeep Mondal and Sourav Mitra for insightful discussions and contributions to the analysis setup. AM thanks Sabine Kraml for her careful reading of the manuscript and constructive comments. AM also acknowledges support from the French Agence Nationale de la Recherche under Grant No. ANR-23-CHRO-0006 (OpenMAPP).

%\newpage
\bibliography{wino_ref}

\newpage

\end{document}